\documentclass[11pt]{article}
\usepackage{amssymb,amsmath}
\usepackage{epsfig}
\usepackage{times,bm,mathrsfs}  
\usepackage{hyperref}

\newtheorem{theorem}{Theorem}[section]

\newtheorem{definition}[theorem]{Definition}

\newtheorem{claim}[theorem]{Claim}
\newtheorem{lemma}[theorem]{Lemma}
\newtheorem{conjecture}[theorem]{Conjecture}

\newtheorem{corollary}[theorem]{Corollary}

\newenvironment{remark}[1][] {
  \smallbreak \noindent {\bf Remark#1:~}} {
  \par\medbreak
}

\newcommand{\qedsymb}{\hfill{\epsfxsize=3mm\epsfbox{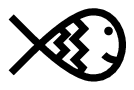}}}
\newenvironment{proof}[1][]{\begin{trivlist}
\item[\hspace{\labelsep}{\bf\noindent Proof#1:\/}]
}{\qedsymb\end{trivlist}}

\def\calF{{\cal F}}

\def\R{\mathbb{R}}

\def\N{\mathbb{N}}

\newcommand\set[2]{\left\{ #1 \left| \; \vphantom{#1 #2} \right. #2  \right\}}
\newcommand\sett[1]{\left\{ #1 \right\}}

\newcommand\defeq{{:=}}
\newcommand\card[1]{\left| #1 \right|}

\newcommand\ip[1]{{\langle {#1} \rangle}}
\newcommand\norm[1]{{\| #1 \|}}

\newcommand{\gauss}[1]{{\tilde{#1}}}
\newcommand{\bunch}[1]{{\overline{#1}}}
\newcommand{\unbunch}[1]{{\underline{#1}}}

\newcommand\ontop[2]{{
\tiny\begin{array}{c} {#1} \\ {#2} \end{array} }}

\newcommand{\onote}[1]{}
\newcommand{\inote}[1]{}
\newcommand{\enote}[1]{}

\newcommand{\eps}{\varepsilon}
\renewcommand{\epsilon}{\varepsilon}

\newcommand{\ignore}[1]{}

\def \E       {{\bf E}}
\def \Var     {{\bf V}}

\def \eps     {\epsilon}
\def \lam     {\lambda}

\def \la      {\langle}
\def \ra      {\rangle}

\def \Reals   {\mathbb{R}}
\def \R       {\Reals}

\def \chop    {\mathrm{chop}}

\def\sat{{\sf sat}}
\def\isat{{\sf isat}}
\def\c{{\psi}}
\def\con{{\Psi}}
\def\four{{\sett{0,1,2,3}}}
\def\three{{\sett{0,1,2}}}
\def\twototwo{{2\hskip -0.3 em \leftrightarrow \hskip -0.3 em 2}}
\def\twotoone{{2\hskip -0.3 em \rightarrow \hskip -0.3 em 1}}
\def\onetoone{{1\hskip -0.3 em \leftrightarrow \hskip -0.3 em 1}}
\def\fish{{\rhd \hskip -0.5 em <}}
\def\Avg     {{A}}

\def\a3c{\textsc{almost-3-coloring}}
\newcommand\col[2]{\textsc{approximate-coloring}({#1},{#2})}

\begin{document}

\title{\bf Conditional Hardness for Approximate Coloring}
\author{
  Irit Dinur\thanks{Hebrew University.  Email: {\tt dinuri@cs.huji.ac.il}.
  Supported by the Israel Science Foundation.}
  \and
  Elchanan Mossel\thanks{Statistics, U.C. Berkeley. Email: {\tt
mossel@stat.berkeley.edu}. Supported by a Miller fellowship in
Computer Science and Statistics and a Sloan fellowship in
Mathematics.}
  \and
  Oded Regev\thanks{Department of Computer Science, Tel-Aviv University, Tel-Aviv 69978, Israel. Supported
   by an Alon Fellowship and by the Israel Science Foundation.}}
\maketitle
\begin{abstract}
We study the $\col q{Q}$ problem: Given a graph $G$, decide
whether $\chi(G) \le q$ or $\chi(G)\ge Q$. We derive {conditional}
hardness for this problem for any constant $3\le q < Q$. For $q\ge
4$, our result is based on Khot's $2$-to-$1$ conjecture [Khot'02].
For $q=3$, we base our hardness result on a certain `$\fish$
shaped' variant of his conjecture.

We also prove that the problem $\a3c_\eps$ is hard for any constant $\eps>0$, assuming Khot's
Unique Games conjecture. This is the problem of
 deciding for a given graph, between the case where
one can $3$-color all but a $\eps$ fraction of the vertices without
monochromatic edges, and the case where the graph contains no
independent set of relative size at least $\eps$.

Our result is based on bounding various generalized noise-stability
quantities using the invariance principle of Mossel et al [MOO'05].
\end{abstract}

\section{Introduction}

For a graph $G=(V,E)$ we let $\chi(G)$ be the chromatic number of
$G$, i.e., the smallest number of colors needed to color the
vertices of $G$ without monochromatic edges. We study the following
problem,

\vskip 3pt $\textbf{\col q Q}:$ Given a graph $G$, decide between
$\chi(G)\le q$ and $\chi(G)\ge Q$. \vskip 3pt

\noindent The problem $\col 3 Q$ is notorious for the wide gap
between the value of $Q$ for which an efficient algorithm is known
and that for which a hardness result exists. The best known
polynomial-time algorithm solves the problem for $Q=\tilde
O(n^{3/14})$ colors, where $n$ is the number of
vertices~\cite{BluKar97}.\footnote{In fact, that algorithm solves
the {\em search problem} of finding a $Q$-coloring given a
$q$-colorable graph. Since we are mainly interested in hardness
results, we restrict our attention to the decision version of the
problem} In contrast, the strongest hardness result shows that the
problem is NP-hard for $Q=5$~\cite{KLS,GuruKhanna}. Thus, the
problem is open for all $5<Q<\tilde O(n^{3/14})$. In this paper we
give some evidence that this problem is hard for any constant
value of $Q$. We remark that any hardness result for $q=3$
immediately carries over for all $q>3$.

The best algorithm known for larger values of $q$ is due to
Halperin et al. \cite{HalperinNatZwick01}, improving on a previous
result of Karger et al \cite{KargerMS98}. Their algorithm solves
$\col q Q$ for $Q=n^{\alpha_q}$ where $0<\alpha_q<1$ is some
function of $q$. For example, $\alpha_4 \approx 0.37$. Improving
on an earlier result of F{\"u}rer \cite{Furer95}, Khot has
shown~\cite{Khot-col} that for any large enough constant $q$ and
$Q=q^{\frac{\log q}{25}}$, $\col q Q$ is NP-hard. Another related
problem is that of approximating the chromatic number
$\chi(\cdot)$ of a given graph. For this problem, an
inapproximability result of $n^{1-o(1)}$ is
known~\cite{FeigeKilian98,Khot-col}.

\paragraph{Constructions:} Our constructions follow the standard composition paradigm initiated
in \cite{BGS,Hastad01}, which has yielded numerous inapproximability
results by now. In our context, this means that
we show reductions from variants of a problem known as label-cover to approximate graph coloring
problems. In the label-cover problem, we are given an undirected
graph and a number $R$. Each edge is associated with a binary relation
on $\{1,\ldots,R\}$. The goal is to label the vertices
with values from $\{1,\ldots,R\}$ such that the number of satisfied edges is maximized,
where an edge is satisfied if its two incident vertices
satisfy the relation associated with it.

As is the case with other composition-based reductions, our reductions
work by replacing each vertex of the label-cover instance with a block
of vertices, known as a {\em gadget}. In other reductions, the gadget is often the binary hypercube $\{0,1\}^R$,
sometimes known as the long-code. In our case, the gadget
is the $q$-ary hypercube, $\{1,\ldots,q\}^R$.
We then connect the gadgets in a way
that ``encodes'' the label-cover constraints. The idea is to ensure
that any $Q$-coloring of the graph (where $Q$ is some constant greater than $q$), can be
``decoded'' into a labeling for the underlying label-cover instance
that satisfies many label-cover constraints.

We note that the idea of using the $q$-ary hypercube as a gadget
has been around for a number of years. This idea has been
studied in \cite{ADFS} and some partial
results were obtained.
The recent progress of \cite{MoOdOl:05} has provided the necessary
tool for achieving our result.

\paragraph{Conjectures:}
Let us now turn our attention to the starting-point label-cover.
None of the known NP-hard label-cover problems (or even more general
PCP systems) seem suitable for composition in our setting. An
increasingly popular approach is to rely on  the `Unique-Games'
conjecture of Khot~\cite{Khot-unique-games}. The conjecture states
that a very restricted version of label-cover is hard. The
strength of this restriction is that in a sense, it reduces the
analysis of the entire construction, to the analysis of the gadget
alone.

However, this conjecture suffers from inherent imperfect
completeness, which prevents it from being used in an approximate
coloring reduction (although it is useful for {\em almost}
approximate coloring). Therefore, we consider restrictions of
label-cover that do have perfect completeness. Our approach is to
search for the least-restricted such label-cover problem that would
still yield the desired result. In all, we consider three
starting-point problems, which result in three different reductions.
\begin{itemize}
\item We show that $\a3c$ is as hard as Khot's Unique Games problem.
\item We show that $\col 4 Q$ is as hard as Khot's
$2$-to-$1$ problem for any constant $Q>0$. This also holds for
$\col q Q$ for any $q\ge 4$.
\item We introduce a new conjecture, which states that label-cover is hard, when the constraints
are restricted to be a certain $\fish$-constraints (read: alpha
constraints). We show that for any constant $Q>3$, $\col 3 Q$ is as hard as solving the
$\fish$-label-cover problem. We remark
that $\fish$-constraints have already appeared in
\cite{DS-VC}.
\end{itemize}

The plausibility of the Unique Games Conjecture, as well as that
of other variants, is uncertain. Very recently,
Trevisan~\cite{Trev05} showed that these conjectures are false
when the parameters are pushed beyond certain sub-constant values.
Hopefully, his result will trigger more attempts to understand
these type of constraint systems from the algorithmic side.

\paragraph{Techniques:}
Our main technique is based on the recent progress of
Mossel et al \cite{MoOdOl:05}. There, they present a powerful technique
for bounding the stability of functions under noise operators.
For example, one special case of their result says that among
all balanced Boolean functions that do not depend too strongly on
any one coordinate, the one that is most stable under noise
is the majority function. In other words, among all such functions,
the majority function is least likely to flip its value if we
flip each of its input bits with some small constant probability.
In fact, this special case was presented as a conjecture in
the work of \cite{KKMO} on $\textsc{MaxCut}$ and
motivated the result of \cite{MoOdOl:05}.

The technique of \cite{MoOdOl:05} is based on what is called
an {\em invariance principle}. This principle allows one to translate
questions in the discrete setting (such as the above question on
the Boolean hypercube) to corresponding questions in other spaces,
and in particular Gaussian space. One then applies known results
in Gaussian space.

In this paper we extend their approach is several respects.
This, we believe, demonstrates the flexibility of the approach
of \cite{MoOdOl:05}.
\begin{itemize}
\item We consider more general noise operators that are given by some
   arbitrary Markov operator. We then apply this to three
   operators, one for each of the aforementioned constructions.
\item We show that when the inner product under noise of two functions
  deviates notably from that of two majority functions, there
  must exist a variable that is influential in {\em both} functions
  (see Theorem \ref{thm:fourier}).
  A direct application of \cite{MoOdOl:05} only yields a variable
  that is influential in {\em one} of the functions. This latter statement
  was enough for the application to $\textsc{MaxCut}$ in \cite{KKMO}.
\item We also present another result tailored for the $\fish$
   constraints (see Theorem \ref{thm:fish_fourier}).
\end{itemize}

\paragraph{Future work:}
Our constructions can be extended in several ways. First, using
similar techniques, one can
show hardness of $\col q Q$ based on the $d$-to-$1$ conjecture of
Khot for larger values of $d$ (and not only $d=2$ as we do here). It would be interesting to
find out how $q$ depends on $d$.
Second, by strengthening the current
conjectures to sub-constant values, one can obtain hardness for
$Q$ that depends on $n$, the number of vertices in the graph.
Again, it is interesting to see how large $Q$ can be. Finally,
let us mention that in all our reductions we in fact
show in the soundness case that there are no independent
sets of relative size larger than $\eps$ for arbitrarily small
constant $\eps$ (note that this
is somewhat stronger than showing that there is no
$Q$-coloring). In fact, a more careful analysis can be used
to obtain the stronger statement that are no `almost-independent'
sets of relative size larger than $\eps$.

\section{Preliminaries}\label{sec:prelim}

\subsection{Approximate coloring problems}
We now define the coloring problems that we study in this paper.
For any $3\le q< Q$, we define {\vskip 5pt \noindent \bf$\col q
Q$:} Given a graph $G$, decide between $\chi(G) \le q$ or
$\chi(G)\ge Q$. \vskip 5pt

For any $\eps>0$ we define
{\vskip 5pt \noindent \bf$\a3c_\eps$:} Given a graph $G=(V,E)$, decide between \begin{itemize}
\item There exists a set $V'\subseteq V$, $\card{V'}\ge
(1-\eps)\card V$ such that $\chi(G|_{V'})=3$ where $G|_{V'}$ is the graph induced by $V'$.
\item Every independent set $S\subseteq V$ in $G$ has size $\card S\le \eps\card V$.
\end{itemize}
Observe that these two items are mutually exclusive for $\eps < 1/4$.

\subsection{Functions on the $q$-ary hypercube}

Let $[q]$ denote the set $\{0,\ldots,q-1\}$. For an element $x$ of
$[q]^n$ write $|x|_a$ for the number of coordinates $k$ of $x$ such
that $x_k =a$ and $|x| = \sum_{a \neq 0} |x|_i$ for the number of
nonzero coordinates.

In this paper we are interested in functions from $[q]^n$ to $\R$.
We define an inner product on this space by $\ip{f,g} =
\frac{1}{q^n} \sum_x f(x)g(x).$ In our applications, we usually take
$q$ to be some constant (say, $3$) and $n$ to be large.

\begin{definition}\label{def:influence}
Let $f : [q]^n \to \R$ be a function. The influence of the $i$'th
variable on $f$, denoted $I_i(f)$ is defined by
\[
I_i(f) = \E[\Var_{x_i}[f(x)|x_1,\ldots,x_{i-1},x_{i+1},\ldots,x_n]]
\]
where $x_1,\ldots,x_n$ are uniformly distributed. 
\end{definition}

Consider a sequence of vectors $\alpha_0 = 1, \alpha_1, \ldots,
\alpha_{q-1} \in \R^q$ forming an orthonormal basis of $\R^q$.
Equivalently, we can think of these vectors as functions from $[q]$
to $\R$. These vectors can be used to form an orthonormal basis of
the space of functions from $[q]^n$ to $\R$, as follows.
\begin{definition}
Let $\alpha_0 = 1, \alpha_1,\ldots,\alpha_{q-1}$ be an orthonormal
basis of $\R^q$. For $x \in [q]^n$, write $\alpha_x$ for
\[
\alpha_{x_1} \otimes \alpha_{x_2} \otimes \cdots \otimes
\alpha_{x_n}.
\]
Equivalently, we can define $\alpha_x$ as the function mapping $y
\in [q]^n$ to $\alpha_{x_1}(y_1) \alpha_{x_2}(y_2) \cdots
\alpha_{x_n}(y_n)$.
\end{definition}

Clearly, any function in $[q]^n \to \R$ can be written as a linear
combination of $\alpha_x$ for $x\in [q]^n$. This leads to the
following definition.

\begin{definition}
For a function $f:[q]^n \to \R$, define $\hat{f}(\alpha_x) = \la
f,\alpha_x \ra$ and notice that $f = \sum_x \hat{f}(\alpha_x)
\alpha_x$.
\end{definition}

\begin{claim}
For any function $f : [q]^n \to \R$ and any $i\in \{1,\ldots,n\}$,
$$ I_i(f) = \sum_{x : x_i \neq 0} \hat{f}^2(\alpha_x).$$
\end{claim}
\begin{proof}
Let us first fix the values of $x_1,\ldots,x_{i-1},x_{i+1},\ldots,x_n$.
Then
\[
\Var_{x_i}[f] = \Var_{x_i}\Big[\sum_y \hat{f}(\alpha_y) \alpha_y\Big] =
\Var_{x_i}\Big[\sum_{y : y_i \neq 0} \hat{f}(\alpha_y) \alpha_y\Big],
\]
where the last equality follows from the fact that if $y_i=0$ then
$\alpha_y$ is a constant function of $x_i$. If $y_i \neq 0$, then the
expected
value of $\alpha_y$ with respect to $x_i$ is zero. Therefore,
\[
\Var_{x_i}\Big[\sum_{y : y_i \neq 0} \hat{f}(\alpha_y) \alpha_y\Big] =
\E_{x_i}
\left[\left(\sum_{y : y_i \neq 0} \hat{f}(\alpha_y) \alpha_y \right)^2
\right] =
\E_{x_i}\left[\sum_{y,z : y_i \neq 0, z_i \neq 0}
\hat{f}(\alpha_y) \hat{f}(\alpha_z) \alpha_y \alpha_z \right].
\]
Thus,
\[
I_i(f) = \E_x \left[\sum_{y,z : y_i \neq 0, z_i \neq 0}
\hat{f}(\alpha_y) \hat{f}(\alpha_z) \alpha_y \alpha_z \right] =
\sum_{y,z : y_i \neq 0, z_i \neq 0}
\hat{f}(\alpha_y) \hat{f}(\alpha_z) \E_x [\alpha_y \alpha_z] =
\sum_{y : y_i \neq 0} \hat{f}^2(\alpha_y),
\]
as needed.
\end{proof}
Notice that this claim holds for any choice of orthonormal basis
$\alpha_0,\ldots,\alpha_{q-1}$ as long as $\alpha_0=1$.

\begin{definition}
Let $f : [q]^n \to \R$ be a function, and let $k\le n$. The
low-level influence of the $i$'th variable on $f$ is defined by
\[
I^{\le k}_i(f) = \sum_{x : x_i \neq 0, \card x \le k}
\hat{f}^2(\alpha_x).
\]
\end{definition}
It is easy to see that for any function $f$, $\sum_i I_i^{\le k}(f) = \sum_{x : \card x \le
k} \hat{f}^2(\alpha_x)\card x \le k \sum_x \hat{f}^2(\alpha_x) =
k\norm{f}_2^2$. In particular, for any function $f$ obtaining values in $[0,1]$,
$\sum_i I_i^{\le k}(f) \le k$.
Moreover, let us mention that $I^{\le k}_i$ is in fact
independent of the particular choice of basis $\alpha_0, \alpha_1,
\ldots,\alpha_{q-1}$ as long as $\alpha_0=1$. This can be verified
from the above definition.

There is a natural equivalence between $[q]^{2n}$ and
$[q^2]^n$. As this equivalence is used often in this paper,
we introduce the following notation.
\begin{definition}\label{def:bunch}
For any $x \in [q]^{2n}$ we denote by $\bunch{x}$ the element of
$[q^2]^n$ given by
$$ \bunch{x} = ((x_1,x_2),\ldots,(x_{2n-1},x_{2n})).$$
For any $y \in [q^2]^n$ we denote by $\unbunch{y}$ the element of $[q]^{2n}$
given by
$$ \unbunch{y} = (y_{1,1}, y_{1,2}, y_{2,1}, y_{2,2}, \ldots, y_{n,1}, y_{n,2}).$$
For a function $f$ on $[q]^{2n}$ we denote by $\bunch{f}$ the function
on $[q^2]^n$ defined by $ \bunch{f}(y) = f(\unbunch{y})$. Similarly,
for a function $f$ on $[q^2]^{n}$ we denote by $\unbunch{f}$ the function
on $[q]^{2n}$ defined by $ \unbunch{f}(x) = f(\bunch{x})$.
\end{definition}

\begin{claim}\label{claim:two-vars}
For any function $f : [q]^{2n} \to \R$, any $i \in \{1,\ldots,n\}$, and any $k \ge 1$,
$$
I^{\le k}_{i}(\bunch{f}) \le I^{\le 2k}_{2i-1}(f) + I^{\le 2k}_{2i}(f).
$$\onote{removing: In particular,
$$
I_{i}(\bunch{f}) \le I_{2i-1}(f) + I_{2i}(f).
$$}
\end{claim}
\begin{proof}
Fix some basis $\alpha_x$ of $[q]^{2n}$
as above and let $\alpha_{\bunch{x}}$ be the basis of $[q^2]^n$ defined by $\alpha_{\bunch{x}}(\bunch{y})=\alpha_x(y)$.
Then, it is easy to see that
$\hat{\bunch{f}}(\alpha_{\bunch{x}}) = \hat{f}(\alpha_{x})$. Hence,
$$ I^{\le k}_{i}(\bunch{f}) =
   \sum_{\bunch{x} : \bunch{x}_i \neq (0,0),|\bunch{x}|\le k }
       \hat{\bunch{f}}^2(\alpha_{\bunch{x}})
\leq \sum_{x : x_{2i-1} \neq 0, |x| \le 2k} \hat{f}^2(\alpha_x) +
     \sum_{x : x_{2i} \neq 0, |x| \le 2k} \hat{f}^2(\alpha_x)
  = I^{\le 2k}_{2i-1}(f) + I^{\le 2k}_{2i}(f)
$$
where we used that $|x| \le 2|\bunch{x}|$.
\end{proof}

For the following definition, recall that we say that a Markov operator
$T$ is {\em symmetric} if it is reversible with respect to the
uniform distribution, i.e., if the transition matrix representing
$T$ is symmetric.

\begin{definition} \label{def:radius}
Let $T$ be a symmetric Markov operator on $[q]$. Let $1 = \lam_0 \geq
\lam_1 \geq \lam_2 \geq \ldots \geq \lam_{q-1}$ be the eigenvalues
of $T$. We define $r(T)$, the {\em spectral radius} of $T$, by
\[
r(T) = \max\{|\lam_1|,|\lam_{q-1}|\}.
\]
\end{definition}

For $T$ as above, we may define a Markov operator $T^{\otimes n}$ on
$[q]^n$ in the standard way. Note that if $T$ is symmetric then
$T^{\otimes n}$ is also symmetric and $r(T^{\otimes n}) = r(T)$. If
we choose $\alpha_0,\ldots,\alpha_{q-1}$ to be an orthonormal set of
eigenvectors for $T$ with corresponding eigenvalues
$\lambda_0,\ldots,\lambda_{q-1}$ (so $\alpha_0=1$), we see that
$$
T^{\otimes n} \alpha_x = \left({\textstyle \prod}_{a \neq 0} \lam_a^{|x|_a} \right) \alpha_x.
$$
and hence
$$
T^{\otimes n} f = \sum_x \left({\textstyle \prod}_{a \neq 0}
\lam_a^{|x|_a}\right) \hat{f}(\alpha_x) \alpha_x.
$$
holds for any function $f:[q]^n \to \R$.

We now describe two operators that we use in this paper.
The first is the Beckner operator, $T_\rho$. For any $\rho \in [-1,1]$, it is defined by $T_\rho(x
\rightarrow x)=\frac{1}{q}+(1-\frac{1}{q})\rho$ and $T_\rho(x \rightarrow
y)=\frac{1}{q}(1-\rho)$ for any $x\neq y$ in $[q]$. It can be seen
that $T_\rho$ is a Markov operator as in Definition \ref{def:radius}
with $\lambda_1=\ldots=\lambda_{q-1}=\rho$ and hence its spectral
radius is $|\rho|$.

Another useful operator is the averaging operator, $\Avg_S$.
For a subset $S \subseteq \{1,\ldots,n\}$, it acts
on functions on $[q]^n$ by averaging over coordinates in $S$, namely,
$$ \Avg_S (f) = \E_{x_S}[f].$$
Notice that the function $\Avg_S(f)$ is independent of the
coordinates in $S$.

\subsection{Functions in Gaussian space}

We let $\gamma$ denote the standard Gaussian measure on $\R^n$. We
denote by $\E_{\gamma}$ the expected value with respect to $\gamma$
and by $\ip{\cdot  ,\cdot }_{\gamma}$ the inner product on
$L^2(\R^n,\gamma)$. Notice that $\E_{\gamma}[f] = \ip{f, {\bf
1}}_\gamma$ where ${\bf 1}$ is the constant $1$ function. For $\rho
\in [-1,1]$, we denote by $U_{\rho}$ the {\em Ornstein-Uhlenbeck
operator}, which acts on $L^2(\R,\gamma)$ by
\[
U_{\rho}f(x) = \E_{y \sim \gamma}[f(\rho x + \sqrt{1-\rho^2} y)].
\]
Finally, for $0 < \mu < 1$, let $F_{\mu} : \R \to \{0,1\}$ denote
the function $F_{\mu}(x) = 1_{x < t}$ where $t$ is chosen in such a
way that $\E_{\gamma}[F_\mu] = \mu$. One useful value that will
appear later is $\ip{F_\eta,U_\rho F_\nu}_\gamma$. For our purposes it is
useful to know that for any $0< \nu, \eta,  -1 \leq \rho \leq 1$, it
holds that
\[
\ip{F_\tau,U_\rho F_\tau}_\gamma \leq \ip{F_\eta,U_\rho F_\nu}_\gamma \leq \tau,
\]
where $\tau = \min(\eta,\nu)$. Moreover, for all $\tau > 0$ and
$\rho > -1$ it holds that
\[
\ip{F_\tau,U_\rho F_\tau}_\gamma > 0.
\]
In fact, it is shown in \cite{RinottRotar:01} that as $\tau \to
0$,
\[
\ip{F_\tau,U_\rho F_\tau}_\gamma \sim \tau^{2/(1+\rho)}
                              (4 \pi \ln(1/\tau))^{-\rho/(1+\rho)}
\frac{(1+\rho)^{3/2}}{(1-\rho)^{1/2}}.
\]
This should play an important role in
possible extensions of our results to cases where $Q$ depends on
$n$.

\section{An Inequality for Noise Operators}

The main result of this section, Theorem \ref{thm:fourier},
is a generalization of the result of \cite{MoOdOl:05}.
It shows that if the inner product of two functions $f$ and $g$ under
some noise operator deviates from a certain range then there must
exist an index $i$ such that the low-level influence of the $i$th
variable is large in both $f$ and $g$. This range depends on the
expected value of $f$ and $g$, and the spectral radius of the
operator $T$.

\begin{theorem} \label{thm:fourier}
Let $q$ be a fixed integer and let $T$ be a symmetric Markov
operator on $[q]$ such that $\rho = r(T) < 1$. Then for any $\eps >
0$ there exist $\delta > 0$ and $k\in \N$ such that if $f,g : [q]^n
\to [0,1]$ are two functions satisfying $\E[f] = \mu, \E[g] = \nu$
and
\begin{equation*}
\min \big( I^{\le k}_i(f),I^{\le k}_i(g) \big) < \delta
\end{equation*}
for all $i$, then it holds that
\begin{equation} \label{eq:lower_inv}
\ip {f, T^{\otimes n} g } \geq
   \ip{ F_{\mu},U_{\rho} (1-F_{1-\nu}) }_{\gamma} - \eps
\end{equation}
and
\begin{equation} \label{eq:upper_inv}
\ip{ f, T^{\otimes n} g} \leq
   \ip{F_{\mu},U_{\rho} F_{\nu}}_{\gamma} + \eps.
\end{equation}
\end{theorem}

Note that (\ref{eq:lower_inv}) follows from (\ref{eq:upper_inv}).
Indeed, apply (\ref{eq:upper_inv}) to $1-g$ to obtain
$$ \ip{ f, T^{\otimes n} (1-g)} \leq
   \ip{F_{\mu},U_{\rho} F_{1-\nu}}_{\gamma} + \eps $$
and then use
$$ \ip{f, T^{\otimes n} (1-g)} = \ip{f,1} - \ip{f, T^{\otimes n} g} =
   \mu - \ip{f, T^{\otimes n} g} = \ip{F_\mu, U_\rho 1}_\gamma - \ip{f, T^{\otimes n} g}.$$
{}From now on we focus on proving (\ref{eq:upper_inv}).

Following the approach of \cite{MoOdOl:05}, the proof consists of
two powerful techniques. The first is an inequality by Christer
Borell \cite{Borell:85} on continuous Gaussian space. The second is
an invariance principle shown in \cite{MoOdOl:05} that allows us to
translate our discrete question to the continuous Gaussian space.

\begin{definition}[Gaussian analogue of an operator] \label{def:gaussian_op}
Let $T$ be an operator as in Definition \ref{def:radius}. We define
its Gaussian analogue as the operator $\gauss{T}$ on
$L^2(\R^{q-1},\gamma)$ given by
\[
\gauss{T} = U_{\lambda_1} \otimes U_{\lambda_2} \otimes \ldots \otimes
U_{\lambda_{q-1}} .
\]
\end{definition}
For example, the Gaussian analogue of $T_\rho$ is $U_\rho^{\otimes
(q-1)}$. We need the following powerful theorem by Borell
\cite{Borell:85}. It says that the functions that maximize the inner
product under the operator $U_\rho$ are the indicator functions of
half-spaces.
\begin{theorem}[Borell~\cite{Borell:85}]\label{thm:borell}
Let $f,g : \R^{n} \to [0,1]$ be two functions and let $\mu =
\E_\gamma[f], \nu = \E_\gamma[g]$. Then
\begin{equation*}
\ip{ f, U_\rho^{\otimes n} g }_{\gamma} \leq \ip{ F_{\mu},U_{\rho} F_{\nu}
}_{\gamma}.
\end{equation*}
\end{theorem}

The above theorem only applies to the Ornstein-Uhlenbeck operator. In the following
corollary we derive a similar statement for more general operators. The proof follows
by writing a general operator as a product of the Ornstein-Uhlenbeck operator
and some other operator.
\begin{corollary} \label{cor:borell_variant}
Let $f,g : \R^{(q-1)n} \to [0,1]$ be two functions satisfying $\E_\gamma[f]
= \mu, \E_\gamma[g] = \nu$. Let $T$ be an operator as in Definition \ref{def:radius} and let $\rho = r(T)$. Then
\begin{equation*}
\la f, \gauss{T}^{\otimes n} g \ra_{\gamma} \leq \la F_{\mu},U_{\rho}
F_{\nu} \ra_{\gamma}.
\end{equation*}
\end{corollary}

\begin{proof}
For $1 \leq i \leq q-1$, let $\delta_i = \lam_i / \rho$. Note that
$|\delta_i| \leq 1$ for all $i$. Let $S$ be the operator defined by
$$ S = U_{\delta_1} \otimes U_{\delta_2} \otimes \ldots \otimes U_{\delta_{q-1}}.$$
Then,
$$U_{\rho}^{\otimes (q-1)} S = U_{\rho} U_{\delta_1} \otimes \ldots \otimes U_{\rho} U_{\delta_{q-1}}
  = U_{\rho \delta_1} \otimes \ldots \otimes U_{\rho \delta_{q-1}}
  = \gauss{T}$$
(this is often called the {\em semi-group property}). It follows
that $\gauss{T}^{\otimes n} = U_{\rho}^{\otimes (q-1)n} S^{\otimes
n}$. The function $S^{\otimes n}g$ obtains values in $[0,1]$ and
satisfies $\E_\gamma[S^{\otimes n}g] = \E_\gamma[g]$. Thus the claim
follows by applying Theorem \ref{thm:borell} to the functions $f$ and
$S^{\otimes n}g$.
\end{proof}

\begin{definition}[Real analogue of a function]
Let $f: [q]^n \to \R$ be a function with decomposition
\[
f = \sum \hat{f}(\alpha_x) \alpha_x.
\]
Consider the $(q-1)n$ variables
$z_1^1,\ldots,z_{q-1}^1,\ldots,z_1^n,\ldots,z_{q-1}^n$ and let
$\Gamma_{x} = \prod_{i=1, x_i \neq 0}^n z_{x_i}^i$. We define the
real analogue of $f$ to be the function $\gauss{f}: \R^{n(q-1)}
\to \R$ given by
\[
\gauss{f} = \sum \hat{f}(\alpha_x) \Gamma_x.
\]
\end{definition}

\begin{claim}\label{clm:discretevscont}
For any two functions $f,g: [q]^n \to \R$ and operator $T$ on
$[q]^n$,
$$ \la f, g \ra = \la \gauss{f}, \gauss{g} \ra_{\gamma} $$
$$ \la f, T^{\otimes n} g \ra = \la \gauss{f}, \gauss{T}^{\otimes n} \gauss{g} \ra_{\gamma} $$
where $\gauss{f},\gauss{g}$ denote the real analogues of
$f,g$ respectively and $\gauss{T}$ denotes the Gaussian analogue of $T$.
\end{claim}
\begin{proof}
Both $\alpha_x$ and $\Gamma_x$ form an orthonormal set of functions
hence both sides of the first equality are
$$ \sum_x \hat{f}(\alpha_x) \hat{g}(\alpha_x).$$
For the second claim, notice that for every $x$, $\alpha_x$ is an eigenvector
of $T^{\otimes n}$ and $\Gamma_x$ is an eigenvector of $\gauss{T}^{\otimes n}$
and both correspond to the eigenvalue ${\textstyle \prod}_{a \neq 0} \lam_a^{|x|_a}$.
Hence, both sides of the second equality are
$$ \sum_x \left( {\textstyle \prod}_{a \neq 0} \lam_a^{|x|_a}\right) \hat{f}(\alpha_x) \hat{g}(\alpha_x) .$$
\end{proof}

\begin{definition}
For any function $f$ with range $\R$, define the function $\chop(f)$
as
$$
\chop(f)(x) = \left\{ \begin{array}{ll}
f(x) & {\rm if ~} f(x) \in [0,1] \\
0    & {\rm if ~} f(x) \leq 0    \\
1    & {\rm if ~} f(x) \geq 1 \end{array} \right.
$$
\end{definition}

The following theorem is proven in \cite{MoOdOl:05}. It shows that
under certain conditions, if a function $f$ obtains values in
$[0,1]$ then $\gauss{f}$ and $\chop(\gauss{f})$ are close. Its proof
is non-trivial and builds on the main technical result of
\cite{MoOdOl:05}, a result that is known as an invariance principal.
In essence, it shows that the distribution of values obtained by $f$
and that obtained by $\gauss{f}$ are close. In particular, since $f$
never deviates from $[0,1]$, it implies that $\gauss{f}$ rarely
deviates from $[0,1]$ and hence $\gauss{f}$ and $\chop(\gauss{f})$
are close. See \cite{MoOdOl:05} for more details.

\begin{theorem}[{\cite[Theorem 3.18]{MoOdOl:05}}]\label{thm:chop} For any $\eta < 1$ and $\eps
> 0$ there exists a $\delta>0$ such that the following holds. For
any function $f:[q]^n \to [0,1]$ such that
$$ \forall x\quad |\hat{f}(\alpha_x)| \leq \eta^{|x|}\qquad and \qquad \forall
i\quad I_i(f) < \delta,$$ then
$$ \|\gauss{f} - \chop(\gauss{f})\| \leq \eps. $$
\end{theorem}

We are now ready to prove the first step in the proof of Theorem
\ref{thm:fourier}. It is here that we use the invariance principle
and Borell's inequality.
\begin{lemma} \label{lem:fourier1}
Let $q$ be a fixed integer and let $T$ be a symmetric Markov
operator on $[q]$ such that $\rho = r(T) < 1$. Then for any $\eps >
0$, $\eta < 1$, there exists a $\delta > 0$ such that for any
functions $f,g : [q]^n \to [0,1]$ satisfying $\E[f] = \mu, \E[g] =
\nu$,
$$ \forall i\qquad \max \left( I_i(f),I_i(g) \right) < \delta $$
and
$$\forall x\qquad |\hat{f}(\alpha_x)| \leq \eta^{|x|}, \quad |\hat{g}(\alpha_x)| \leq
\eta^{|x|}, $$ it holds that
\begin{equation*}
\ip{ f, T^{\otimes n} g} \leq
   \ip{F_{\mu},U_{\rho} F_{\nu}}_{\gamma} + \eps.
\end{equation*}
\end{lemma}
\begin{proof}
Let $\mu' = \E_{\gamma}[\chop(\gauss{f})]$ and $\nu' =
\E_{\gamma}[\chop(\gauss{g})]$. We note that $\la F_{\mu}, U_{\rho}
F_{\nu} \ra_\gamma$ is a uniformly continuous function of $\mu$ and $\nu$.
Let $\eps_1$ be chosen such that if $|\mu-\mu'| \leq \eps_1$ and
$|\nu-\nu'| \leq \eps_1$ then it holds that
\[
|\la F_{\mu}, U_{\rho} F_{\nu} \ra_\gamma - \la F_{\mu'}, U_{\rho} F_{\nu'}
\ra_\gamma| \leq \eps/2.
\]
Let $\eps_2 = \min(\eps/4, \eps_1)$ and let $\delta = \delta(\eta,
\eps_2)$ be the value given by Theorem~\ref{thm:chop} with $\eps$
taken to be $\eps_2$. Then, using the Cauchy-Schwartz inequality,
$$ |\mu' - \mu| = |\E_{\gamma}[\chop(\gauss{f}) - \gauss{f} ]| = |\ip{\chop(\gauss{f}) - \gauss{f}, {\bf 1}}_\gamma| \le
\|\chop(\gauss{f})-\gauss{f}\| \le \eps_2 \le \eps_1.$$
Similarly, we have $|\nu'-\nu| \leq \eps_1$. Now,
\begin{align*}
\la f, T^{\otimes n} g \ra
 &= \la \gauss{f}, \gauss{T}^{\otimes n} \gauss{g} \ra_{\gamma} && \text{(Claim~\ref{clm:discretevscont})}\\
 &= \ip{ \chop(\gauss{f}), \gauss{T}^{\otimes n} \chop(\gauss{g}) }_{\gamma} + \\
       &\qquad \la \chop(\gauss{f}), \gauss{T}^{\otimes n} (\gauss{g}-\chop(\gauss{g})) \ra_{\gamma} +
       \la \gauss{f} - \chop(\gauss{f}) , \gauss{T}^{\otimes n} \gauss{g} \ra_{\gamma}\\
 &\leq \la \chop(\gauss{f}), \gauss{T}^{\otimes n} \chop(\gauss{g}) \ra_{\gamma} + 2 \eps_2 \\
 &\leq \la F_{\mu'}, U_{\rho} F_{\nu'} \ra_{\gamma} + 2 \eps_2 &&\text{(Corollary \ref{cor:borell_variant})}\\
 &\leq \la F_{\mu}, U_{\rho} F_{\nu} \ra_{\gamma} + \eps/2 + 2 \eps_2
  \leq \la F_{\mu}, U_{\rho} F_{\nu} \ra_{\gamma} + \eps
\end{align*}
where the first inequality follows from the Cauchy-Schwartz inequality
together with the fact that $\chop(\gauss{f})$ and $\gauss{g}$ have
$L_2$ norm at most $1$ and that $\gauss{T}^{\otimes n}$ is a
contraction on $L_2$.
\end{proof}

We complete the proof of Theorem \ref{thm:fourier} by proving:
\begin{lemma} \label{lem:min_max_inf}
Let $q$ be a fixed integer and let $T$ be a symmetric Markov
operator on $[q]$ such that $\rho = r(T) < 1$. Then for any $\eps >
0$, there exists a $\delta > 0$ and an integer $k$ such that for any
functions $f,g : [q]^n \to [0,1]$ satisfying $\E[f] = \mu, \E[g] =
\nu$ and
\begin{equation} \label{eq:min_inf}
\forall i\quad\min \big( I_i^{\leq k}(f),I_i^{\leq k}(g) \big) <
\delta
\end{equation}
then
\begin{equation} \label{eq:fg_bd}
\ip{ f, T^{\otimes n} g} \leq
   \ip{F_{\mu},U_{\rho} F_{\nu}}_{\gamma} + \eps.
\end{equation}
\end{lemma}

\begin{proof}
Let $f_1 = T_{\eta}^{\otimes n} f$ and $g_1 = T_{\eta}^{\otimes n} g$ where $\eta < 1$ is
chosen so that $\rho^j(1-\eta^{2j}) < \eps/4$ for all $j$. Then
\begin{align*}
|\la f_1, T^{\otimes n} g_1 \ra - \la f, T^{\otimes n} g \ra| &=
 \Big|\sum_{x}
 \hat{f}(\alpha_x) \hat{g}(\alpha_x) \prod_{a \neq 0} \lam_a^{|x|_a}
 (1-\eta^{2|x|}) \Big| \\
&\leq
  \sum_{x} \rho^{|x|} (1-\eta^{2|x|})
  \Big| \hat{f}(\alpha_x) \hat{g}(\alpha_x) \Big|
  \leq \eps/4
\end{align*}
where the last inequality follows from the Cauchy-Schwartz inequality.
Thus, in order to prove~(\ref{eq:fg_bd}) it suffices to prove
\begin{equation} \label{eq:fg_tilde}
\ip{ f_1, T^{\otimes n} g_1} \leq
   \ip{F_{\mu},U_{\rho} F_{\nu}}_{\gamma} + 3\eps/4.
\end{equation}

Let $\delta(\eps/4, \eta)$ be the value given by
Lemma~\ref{lem:fourier1} plugging in $\eps/4$ for $\eps$. Let
$\delta' = \delta(\eps/4, \eta)/2$. Let $k$ be chosen so that
$\eta^{2k} < \min(\delta',\eps/4)$. Define $C = k/\delta'$ and
$\delta = (\eps / 8C)^2 < \delta'$ . Let
\[
B_f = \{i : I_i^{\leq k}(f) \geq \delta'\}, \quad B_g = \{i :
I_i^{\leq k}(g) \geq \delta'\}.
\]
We note that $B_f$ and $B_g$ are of size at most $C = k/\delta'$.
By \eqref{eq:min_inf}, we have that whenever $i\in B_f$, $I_i^{\leq
k}(g) < \delta$.
Similarly, for every $i \in B_g$ we have $I_i^{\leq k}(f) <
\delta$.
In particular, $B_f$ and $B_g$ are disjoint.

Recall the averaging operator $\Avg$. We now let
\begin{align*}
f_2(x) &= \Avg_{B_f}(f_1) =
\sum_{x : x_{B_f} = 0} \hat{f}(\alpha_x)\alpha_x \eta^{|x|},  \\
g_2(x) &= \Avg_{B_g}(g_1) = \sum_{x : x_{B_g} = 0}
\hat{g}(\alpha_x)\alpha_x \eta^{|x|}.
\end{align*}
Clearly, $\E[f_2] = \E[f]$ and $\E[g_2] = \E[g]$, and for all $x$
$f_2(x),g_2(x)\in [0,1]$. It is easy to see that $I_i(f_2) = 0$ if
$i \in B_f$ and $I_i(f_2) \le I^{\leq k}_i(f) + \eta^{2k} < 2
\delta'$ otherwise and similarly for $g_2$. Thus, for any $i$, $\max
\left( I_i(f_2),I_i(g_2) \right) < 2 \delta'$. We also see that for
any $x$, $|\hat{f_2}(\alpha_x)| \le \eta^{|x|}$ and the same for
$g_2$. Thus, we can apply Lemma~\ref{lem:fourier1} to obtain that
\[
\la f_2, T^{\otimes n} g_2 \ra \leq \ip{F_{\mu},U_{\rho}
F_{\nu}}_{\gamma} + \eps/4.
\]

In order to show (\ref{eq:fg_tilde}) and complete the proof, we show
that
\[
|\la f_1, T^{\otimes n} g_1 \ra - \la f_2, T^{\otimes n}
g_2 \ra| \le \eps/2.
\]
This follows by
\begin{align*}
|\la f_1, T^{\otimes n} g_1 \ra - \la f_2, T^{\otimes n}
g_2 \ra| &=
 \Big|\sum_{x : x_{B_f\cup B_g} \neq 0}
 \hat{f}(\alpha_x) \hat{g}(\alpha_x) \prod_{a \neq 0} \lam_a^{|x|_a} \eta^{2|x|}
\Big| \\
&\leq  \eta^{2k} \sum_{x : |x| \ge k}
  \Big| \hat{f}(\alpha_x) \hat{g}(\alpha_x) \Big| +
  \sum \left\{ \Big| \hat{f}(\alpha_x) \hat{g}(\alpha_x) \Big| :
  x_{B_f\cup B_g} \neq 0, |x| \leq k \right\}
\\
&\leq  \eps/4 + \sum_{i \in B_f \cup B_g}
\sum \left\{\Big| \hat{f}(\alpha_x) \hat{g}(\alpha_x) \Big| : x_i \neq 0, |x| \leq k \right\}\\
 &\leq \eps/4 + \sum_{i \in B_f \cup B_g} \sqrt{I^{\leq k}_i(f)} \sqrt{I^{\leq k}_i(g)} \\
 &\leq \eps/4 + \sqrt{\delta} (|B_f| + |B_g|) \\
 & \leq \eps/4+ 2 C \sqrt{\delta} = \eps/2,
\end{align*}
where the next-to-last inequality holds because for each $i \in B_f \cup B_g$
one of $I^{\leq k}_i(f), I^{\leq k}_i(g)$ is at most $\delta$ and
the other is at most $1$.
\end{proof}

The final theorem of this section is needed only for the $\col 3
Q$ result. Here, the operator $T$ acts on $[q^2]$ and is assumed
to have an additional property. Before proceeding, it is helpful
to recall Definition \ref{def:bunch}.

\begin{theorem} \label{thm:fish_fourier}
Let $q$ be a fixed integer and let $T$ be a symmetric Markov
operator on $[q^2]$ such that $\rho = r(T) < 1$. Suppose moreover,
that $T$ has the following property. Given $(x_1,x_2)$ chosen uniformly
at random and $(y_1,y_2)$ chosen according to $T$ applied to $(x_1,x_2)$
we have that $(x_2,y_2)$ is distributed uniformly at random.
Then for any $\eps > 0$, there exists a $\delta > 0$ and an integer
$k$ such that for any functions $f,g : [q]^{2n} \to [0,1]$ satisfying
$\E[f] = \mu, \E[g] = \nu$, and for $i=1,\ldots,n$
\begin{equation*}
\min \big( I_{2i-1}^{\leq k}(f),I_{2i-1}^{\leq k}(g) \big) < \delta,
\qquad
\min \big( I_{2i-1}^{\leq k}(f),I_{2i}^{\leq k}(g) \big) < \delta, \text{ and}
\qquad
\min \big( I_{2i}^{\leq k}(f),I_{2i-1}^{\leq k}(g) \big) < \delta
\end{equation*}
it holds that
\begin{equation} \label{eq:lower_inv_fish}
\ip{ \bunch{f}, T^{\otimes n} \bunch{g}}  \geq
   \ip{ F_{\mu},U_{\rho} (1-F_{1-\nu}) }_{\gamma} - \eps
\end{equation}
and
\begin{equation} \label{eq:upper_inv_fish}
\ip{ \bunch{f}, T^{\otimes n} \bunch{g}} \leq
   \ip{F_{\mu},U_{\rho} F_{\nu}}_{\gamma} + \eps.
\end{equation}

\end{theorem}
\begin{proof}
As in Theorem \ref{thm:fourier}, \eqref{eq:lower_inv_fish} follows from \eqref{eq:upper_inv_fish}
so it is enough to prove \eqref{eq:upper_inv_fish}.
Assume first that in addition to the three conditions above we also have that
for all $i=1,\ldots,n$,
\begin{equation}\label{eq:extra_condition}
\min \big( I_{2i}^{\leq k}(f),I_{2i}^{\leq k}(g) \big) < \delta.
\end{equation}
Then it follows that for all $i$,
either both $I_{2i-1}^{\leq k}(f)$ and $I_{2i}^{\leq k}(f)$ are smaller than $\delta$
or both $I_{2i-1}^{\leq k}(g)$ and $I_{2i}^{\leq k}(g)$ are smaller than $\delta$.
Hence, by Claim \ref{claim:two-vars}, we know that for all $i$ we have
\[
\min \left( I_{i}^{\leq k/2}(\bunch{f}),I_{i}^{\leq k/2}(\bunch{g}) \right) < 2\delta
\]
and the result then follows from Lemma \ref{lem:min_max_inf}.
However, we do not have this extra condition and hence we have to
deal with `bad' coordinates $i$ for which $\min(I_{2i}^{\leq
k}(f),I_{2i}^{\leq k}(g)) \ge \delta$. Notice for such $i$ it must
be the case that both $I_{2i-1}^{\leq k}(f)$ and $I_{2i-1}^{\leq
k}(g)$ are smaller than $\delta$. Informally, the proof proceeds as
follows. We first define functions $f_1,g_1$ that are obtained from
$f,g$ by adding a small amount of noise. We then obtain $f_2,g_2$
from $f_1,g_1$ by averaging the coordinates $2i-1$ for bad $i$.
Finally, we obtain $f_3,g_3$ from $f_2,g_2$ by averaging the
coordinate $2i$ for bad $i$. The point here is to maintain
$\ip{\bunch f,T ^{\otimes n}\bunch g} \approx \ip{\bunch
f_1,T^{\otimes n} \bunch g_1} \approx \ip{\bunch f_2,T^{\otimes n}
\bunch g_2} \approx \ip{\bunch f_3,T^{\otimes n}\bunch g_3}$. The
condition in Equation \ref{eq:extra_condition} now applies to
$f_3,g_3$ and we can apply Lemma \ref{lem:min_max_inf}, as described
above. We now describe the proof in more detail.

We first define $\bunch{f_1} = T_{\eta}^{\otimes n} \bunch{f}$ and $\bunch{g_1} = T_{\eta}^{\otimes n} \bunch{g}$
where $\eta < 1$ is chosen so that $\rho^{j}(1-\eta^{2j}) < \eps/4$ for all $j$. As in
the previous lemma it is easy to see that
\[
|\la \bunch{f_1}, T^{\otimes n} \bunch{g_1} \ra - \la \bunch{f}, T^{\otimes n} \bunch{g} \ra| <
\eps/4
\]
and thus it suffices to prove that
\[
\ip{ \bunch{f_1}, T^{\otimes n} \bunch{g_1}} \leq
   \ip{F_{\mu},U_{\rho} F_{\nu}}_{\gamma} + 3\eps/4.
\]

Let $\delta(\eps/2, \eta),k(\eps/2,\eta)$ be the values given by Lemma \ref{lem:min_max_inf}
with $\eps$ taken to be $\eps/2$. Let $\delta'=\delta(\eps/2,\eta)/2$.
Choose a large enough $k$ so that $128 k \eta^{k} <
\eps^2 \delta'$ and $k/2 > k(\eps/2,\eta)$. We let $C = k/\delta'$ and $\delta = \eps^2 / 128C$.
Notice that $\delta < \delta'$ and $\eta^{k} < \delta$. Finally, let
\[
B = \set{~i~}{~I_{2i}^{\leq k}(f) \geq \delta', I_{2i}^{\leq k}(g) \geq
\delta'}.
\]
We note that $B$ is of size at most $C$. We also note that if $i \in
B$ then we have $I_{2i-1}^{\leq k}(f) < \delta$ and $I_{2i-1}^{\leq k}(g) < \delta$.
We claim that this implies that $I_{2i-1}(f_1) \leq \delta + \eta^{k} < 2 \delta$ and similarly
for $g$. To see that, take any orthonormal basis $\beta_0=1,\beta_1,\ldots,\beta_{q-1}$ of $\R^q$
and notice that we can write
$$ f_1 = \sum_{x \in [q]^{2n}} \hat{f}(\beta_x) \eta^{|\bunch{x}|} \beta_x .$$
Hence,
$$ I_{2i-1}(f_1) =
   \sum_{\ontop{x \in [q]^{2n}}{ x_{2i-1} \neq 0}} \hat{f}(\beta_x)^2 \eta^{2|\bunch{x}|} <
   \delta + \eta^{k} \sum_{\ontop{x \in [q]^{2n}}{ |x| > k}} \hat{f}(\beta_x)^2 \le \delta + \eta^{k}$$
where we used that the number of nonzero elements in $\bunch{x}$ is at least
half of that in $x$.

Next, we define $f_2 = \Avg_{2B-1}(f_1)$ and $g_2 = \Avg_{2B-1}(g_1)$
where $\Avg$ is the averaging operator and $2B-1$ denotes the set $\set{2i-1}{i \in B}$.
Note that
\[
\| \bunch{f_2} - \bunch{f_1} \|_2^2 = \| f_2-f_1 \|_2^2 \leq \sum_{i \in B} I_{2i-1}(f_1) \leq 2 C \delta.
\]
and similarly,
\[
\| \bunch{g_2} - \bunch{g_1} \|_2^2 = \| g_2-g_1 \|_2^2 \leq 2 C \delta.
\]
Thus
\begin{align*}
 | \ip{ \bunch{f_1}, T^{\otimes n} \bunch{g_1}} - \ip{ \bunch{f_2}, T^{\otimes n} \bunch{g_2}} |
&\leq | \ip{ \bunch{f_1}, T^{\otimes n} \bunch{g_1}} - \ip{ \bunch{f_1}, T^{\otimes n} \bunch{g_2}}
| + | \ip{ \bunch{f_1}, T^{\otimes n} \bunch{g_2}} - \ip{ \bunch{f_2}, T^{\otimes n} \bunch{g_2}}
| \\
&\leq 2 \sqrt{2C \delta} = \eps/4
\end{align*}
where the last inequality follows from the
Cauchy-Schwartz inequality together with the fact that $\norm{\bunch
{f_1}}_2 \le 1$ and also $\norm{{T}^{\otimes n}\bunch {g_2}}_2\le 1$. Hence, it suffices
to prove
$$
\ip{ \bunch{f_2}, T^{\otimes n} \bunch{g_2}} \leq
   \ip{F_{\mu},U_{\rho} F_{\nu}}_{\gamma} + \eps/2.
$$

We now define $f_3 = \Avg_{2B}(f_2)$ and $g_3 = \Avg_{2B}(g_2)$. Equivalently, we have
$\bunch{f_3} = \Avg_B(\bunch{f_1})$ and $\bunch{g_3} = \Avg_B(\bunch{g_1})$. We
 show that $\ip{ \bunch{f_2}, T^{\otimes n} \bunch{g_2}} = \ip{ \bunch{f_3}, T^{\otimes n} \bunch{g_3}}$.
Let $\alpha_x$, $x \in [q^2]^n$, be an orthonormal basis of eigenvectors of $T^{\otimes n}$. Then
\begin{align*}
\ip{ \bunch{f_3}, T^{\otimes n} \bunch{g_3}} &= \sum_{x,y \in [q^2]^n, x_B=y_B=0}
    \hat{\bunch{f_1}}(\alpha_x)\hat{\bunch{g_1}}(\alpha_y) \ip{\alpha_x, T^{\otimes n} \alpha_y}.
\end{align*}
Moreover, since $\Avg$ is a linear operator and $f_1$ can be written as
$\sum_{x\in [q^2]^n} \hat{\bunch{f_1}}(\alpha_x)\unbunch{\alpha_x}$ and similarly for $g_1$, we have
\begin{align*}
\ip{ \bunch{f_2}, T^{\otimes n} \bunch{g_2}} &= \sum_{x,y \in [q^2]^n}
    \hat{\bunch{f_1}}(\alpha_x)\hat{\bunch{g_1}}(\alpha_y) \ip{\bunch{\Avg_{2B-1}(\unbunch{\alpha_x})},
          T^{\otimes n} \bunch{\Avg_{2B-1}(\unbunch{\alpha_y})}}.
\end{align*}
First, notice that when $x_B=0$, $\bunch{\Avg_{2B-1}(\unbunch{\alpha_x})} = \alpha_x$ since
$\alpha_x$ does not depend on coordinates in $B$. Hence, in order to show that the two expressions
above are equal, it suffices to show that
$$\ip{\bunch{\Avg_{2B-1}(\unbunch{\alpha_x})}, T^{\otimes n} \bunch{\Avg_{2B-1}(\unbunch{\alpha_y})}}=0$$
unless $x_B=y_B=0$. So assume without loss of generality that $i\in B$ is such that $x_i \neq 0$.
The above inner product can be equivalently written as
$$ \E_{z, z' \in [q^2]^n} [\bunch{\Avg_{2B-1}(\unbunch{\alpha_x})}(z) \cdot \bunch{\Avg_{2B-1}(\unbunch{\alpha_y})}(z')] $$
where $z$ is chosen uniformly at random and $z'$ is chosen according to $T^{\otimes n}$ applied to $z$.
Fix some arbitrary values to $z_1,\ldots,z_{i-1},z_{i+1},\ldots,z_n$ and $z'_1,\ldots,z'_{i-1},z'_{i+1},\ldots,z'_n$
and let us show that
$$ \E_{z_i, z'_i \in [q^2]} [\bunch{\Avg_{2B-1}(\unbunch{\alpha_x})}(z) \cdot \bunch{\Avg_{2B-1}(\unbunch{\alpha_y})}(z')] = 0.$$
Since $i\in B$, the two expressions inside the expectation do not depend on $z_{i,1}$
and $z'_{i,1}$ (where by $z_{i,1}$ we mean the first coordinate of $\unbunch{z_i}$). Moreover, by our assumption on $T$, $z_{i,2}$ and
$z'_{i,2}$ are independent. Hence, the above
expectation is equal to
$$ \E_{z_i \in [q^2]} [\bunch{\Avg_{2B-1}(\unbunch{\alpha_x})}(z)] \cdot
   \E_{z'_i \in [q^2]} [\bunch{\Avg_{2B-1}(\unbunch{\alpha_y})}(z')].$$
Since $x_i \neq 0$, the first expectation is zero.
This establishes that $\ip{ \bunch{f_2}, T^{\otimes n} \bunch{g_2}} = \ip{ \bunch{f_3}, T^{\otimes n} \bunch{g_3}}$.

The functions $f_3,g_3$ satisfy the property that for every $i=1,\ldots,n$, either
both $I^{\le k}_{2i-1}(f_3)$ and $I^{\le k}_{2i}(f_3)$ are smaller than $\delta'$
or both $I^{\le k}_{2i-1}(g_3)$ and $I^{\le k}_{2i}(g_3)$ are smaller than $\delta'$.
By Claim \ref{claim:two-vars}, we get that for $i=1,\ldots,n$, either $I^{\le k/2}_{i}(\bunch{f_3})$
or $I^{\le k/2}_{i}(\bunch{g_3})$ is smaller $2\delta'$. We can now apply Lemma \ref{lem:min_max_inf}
to obtain
\[
\ip{ \bunch{f_3}, T^{\otimes n} \bunch{g_3}} \leq
   \ip{F_{\mu},U_{\rho} F_{\nu}}_{\gamma} + \eps/2.
\]
\end{proof}

\section{Approximate Coloring}

In this section we describe and prove reductions to the three
problems described in Section~\ref{sec:prelim}, based on three
conjectures on the hardness of label-cover. These conjectures, along
with some definitions, are described in
Section~\ref{subsec:labelcover}. The three reductions are very
similar, each combining a conjecture with an appropriately
constructed noise operator. In Section~\ref{subsec:noise} we
describe the three noise operators, and in
Section~\ref{subsec:constr} we spell out the constructions. Then, in
Sections~\ref{subsec:complete} and \ref{subsec:sound}, we prove the
completeness and soundness of the three reductions.

\subsection{Label-cover problems}\label{subsec:labelcover}

\begin{definition}
A label-cover instance is a triple $G=((V,E),R,\con)$
where $(V,E)$ is a graph, $R$ is an integer,
and $\con = \set{\c_e \subseteq \sett{1,\ldots,R}^2}{e\in
E}$ is a set of constraints (relations), one for each edge. For a given
labeling $L:V\to\sett{1,\ldots,R}$, let
$$\sat_L(G) = \Pr_{e=(u,v)\in E}[(L(u),L(v))\in\c_e],\qquad\sat(G) =
\max_{L}(\sat_L(G))\,.$$
\end{definition}

For $t,R\in\mathbb N$ let $R \choose {\le t}$ denote the
collection of all subsets of $\sett{1,\ldots,R}$ whose size is at
most $t$.

\begin{definition}\label{def:multi-labeling}
A $t$-labeling is a function $L:V\to {R\choose {\le t}}$ that
labels each vertex $v\in V$ with a subset of values $L(v) \subseteq
\sett{1,\ldots,R}$ such that $\card{L(v)} \le t$ for all $v\in V$. A
$t$-labeling $L$ is said to {\em satisfy} a constraint $\c\subseteq
\sett{1,\ldots,R}^2$ over variables $u$ and $v$ iff there are $a\in
L(u)$, $b\in L(v)$ such that $(a,b)\in \c$. In other words, iff
$(L(u)\times L(v)) \cap \c \neq \phi$.
\end{definition}
For the special case of $t=1$, a $1$-labeling is simply a labeling
$L:V\to \sett{1,\ldots,R}$. In this case, a constraint $\c$ over
$u,v$ is satisfied by $L$ iff $(L(u),L(v))\in\c$.

Similar to the definition of $\sat(G)$, we also define $\isat(G)$
(``induced-$\sat$'') to be the relative size of the largest set of
vertices for which there is a labeling that satisfies {\em all} of
the induced edges.
$$ \isat(G) = \max_S{\set{\frac{\card S}{\card V}}{\exists L:S\to\sett{1,\ldots,R}
\hbox{
that satisfies all the constraints induced by }S\subseteq V}}.$$
Let $\isat_t(G)$ denote the relative size of the largest set of
vertices $S\subseteq V$ for which there is a $t$-labeling that
satisfies {\em all} the constraints induced by $S$.
$$ \isat_t(G) =
\max_S{\set{\frac {\card S}{\card V}}{\exists L:S\to{R\choose
{\le t}}\hbox{ that satisfies all the constraints induced by
}S\subseteq V}}.$$

\begin{figure}[t]
\center{\epsfbox{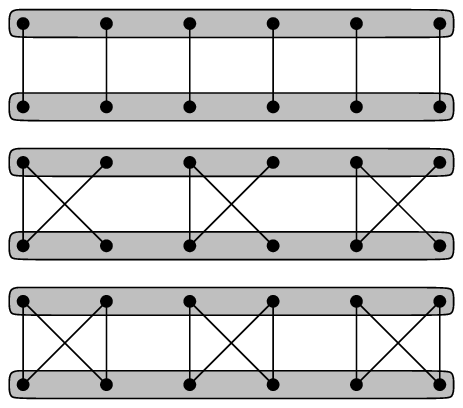}}
 \caption{Three types of constraints (top to bottom): $\onetoone, \fish, \twototwo$}
 \label{fig:constraints}
\end{figure}

We next describe three conjectures on which our reductions are
based. The main difference between the three conjectures is in the type of
constraints that are allowed. The three types are defined next, and
also illustrated in Figure~\ref{fig:constraints}.

\begin{definition}[$\onetoone$-constraint]\label{def:1-to-1} A
$\onetoone$ constraint is a relation $\sett{(i,\pi(i))}_{i=1}^R$,
where $\pi:\sett{1,\ldots,R}\to\sett{1,\ldots,R}$ is any arbitrary
permutation. The constraint is satisfied by $(a,b)$ iff $b=\pi(a)$.
\end{definition}

\begin{definition}[$\twototwo$-constraint]\label{def:2-to-2} A
$\twototwo$ constraint is defined by a pair of permutations
$\pi_1,\pi_2:\sett{1,\ldots,2R}\to\sett{1,\ldots,2R}$ and the
relation $$\twototwo =
\sett{(2i,2i),(2i,2i-1),(2i-1,2i),(2i-1,2i-1)}_{i=1}^R\,.$$ The
constraint is satisfied by $(a,b)$ iff
$(\pi_1^{-1}(a),\pi^{-1}_2(b)) \in \twototwo$.
\end{definition}

\begin{definition}[$\fish$-constraint]\label{def:alpha}
An $\fish$ constraint is defined by a pair of permutations
$\pi_1,\pi_2:\sett{1,\ldots,2R}\to\sett{1,\ldots,2R}$ and the
relation $$\fish =
\sett{(2i-1,2i-1),(2i,2i-1),(2i-1,2i)}_{i=1}^R\,.$$ The constraint
is satisfied by $(a,b)$ iff $(\pi^{-1}_1(a),\pi^{-1}_2(b)) \in
\fish$.
\end{definition}

\begin{conjecture}[$\onetoone$ Conjecture]\label{conj:one-to-one} For every $\eps,\zeta>0$ and
$t\in\mathbb{N}$ there exists some $R\in\mathbb{N}$ such that given
a label-cover instance $G=\ip{(V,E),R,\con}$ where
all constraints are $\onetoone$-constraints, it is NP-hard to decide
between
\begin{itemize}
\item $\isat(G) \ge 1-\zeta$
\item $\isat_t(G) < \eps$
\end{itemize}
\end{conjecture}
It is easy to see that the above problem is in P when $\zeta=0$.

\begin{conjecture}[$\twototwo$ Conjecture]\label{conj:two-to-two} For every $\eps>0$ and
$t\in\mathbb{N}$ there exists some $R\in\mathbb{N}$ such that given
a label-cover instance $G=\ip{(V,E),2R,\con}$ where
all constraints are $\twototwo$-constraints, it is NP-hard to decide
between
\begin{itemize}
\item $\sat(G) = 1$
\item $\isat_t(G) < \eps$
\end{itemize}
\end{conjecture}

The above two conjectures are no stronger than the corresponding
conjectures of Khot. Namely, our $\onetoone$ conjecture is not
stronger than Khot's (bipartite) unique games conjecture, and our
$\twototwo$ conjecture is not stronger than Khot's (bipartite)
$\twotoone$ conjecture. The former claim was already proven by
Khot and Regev in \cite{KhotR03}. The latter claim is proven in a
similar way. For completeness, we include both proofs in
Appendix~\ref{app:Conjectures}. We also make a third conjecture
that is used in our reduction to $\col 3 Q$. This conjecture seems
stronger than Khot's conjectures.

\begin{conjecture}[$\fish$ Conjecture]\label{conj:alpha} For every $\eps>0$ and
$t\in\mathbb{N}$ there exists some $R\in\mathbb{N}$ such that given
a label-cover instance $G=\ip{(V,E),2R,\con}$ where
all constraints are $\fish$-constraints, it is NP-hard to decide
between
\begin{itemize}
\item $\sat(G) = 1$
\item $\isat_t(G) < \eps$
\end{itemize}
\end{conjecture}

\begin{remark}
The (strange-looking) $\fish$-shaped constraints have already
appeared before, in \cite{DS-VC}. There, it is (implicitly) proven
that for all $\eps,\zeta>0$ given a label-cover instance $G$ where
all constraints are $\fish$-constraints, it is NP-hard to
distinguish between
\begin{itemize}
\item $\isat(G)>1-\zeta$
\item $\isat_{t=1}(G)<\eps$
\end{itemize}
The main difference between their case and our conjecture is
that in our conjecture we consider any constant $t$, while in
their case $t$ is $1$.
\end{remark}

\subsection{Noise operators}\label{subsec:noise}
\begin{figure}[h]
\begin{center}
\begin{tabular}{ccc}
\epsfbox{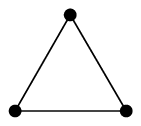}  & \epsfbox{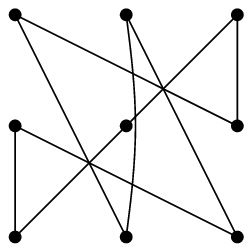} & \epsfbox{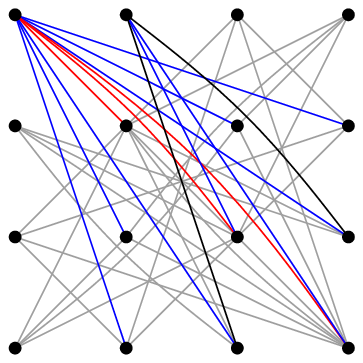} \\
(a) & (b) & (c)
\end{tabular}
 \caption{Three noise operators (edge weights not shown) corresponding
 to: (a) $\a3c$, (b) $\col 3 Q$ and (c) $\col 4 Q$.}
 \label{fig:noiseoperators}
\end{center}
\end{figure}

We now define the noise operators corresponding to the
$\onetoone$-constraints, $\fish$-constraints, and
$\twototwo$-constraints.
The noise operator that corresponds to the
$\onetoone$-constraints is the simplest, and acts on $\three$.  For
the other two cases, since the constraints involve pairs of
coordinates, we obtain
an operator on $\three^2$ and an operator on $\four^2$.
See Figure \ref{fig:noiseoperators} for an illustration.

\begin{lemma}\label{lemma:noise-almost}
There exists a symmetric Markov operator $T$ on $\three$ such that
$r(T) < 1$ and such that if $T(x \leftrightarrow y) > 0$ then $x \neq y$.
\end{lemma}
\begin{proof}
Take the operator given by
$$ T =
\left(
\begin{matrix} 0 & 1/2 & 1/2 \\
               1/2 & 0 & 1/2 \\
               1/2 & 1/2 & 0
\end{matrix} \right).
$$
See Figure \ref{fig:noiseoperators}(a).
\end{proof}

\begin{lemma}\label{lemma:noise-2by2}
There exists a symmetric Markov operator $T$ on $\four^2$ such that
$r(T) < 1$ and such that if $T((x_1,x_2) \leftrightarrow (y_1,y_2)) > 0$ then
$\{x_1,x_2\} \cap \{y_1,y_2\} = \emptyset$.
\end{lemma}

\begin{proof}
Our operator has three types of
transitions, with transitions probabilities $\beta_1,\beta_2$, and $\beta_3$.
\begin{itemize}
\item
With probability $\beta_1$ we have $(x,x) \leftrightarrow (y,y)$ where $x \neq
y$.
\item
With probability $\beta_2$ we have $(x,x) \leftrightarrow (y,z)$ where $x,y,z$ are all different.
\item
With probability $\beta_3$ we have $(x,y) \leftrightarrow (z,w)$ where $x,y,z,w$ are all
different.
\end{itemize}
These transitions are illustrated in Figure \ref{fig:noiseoperators}(c)
with red indicating $\beta_1$ transitions, blue indicating $\beta_2$ transitions,
and black indicating $\beta_3$ transitions.
For $T$ to be symmetric Markov operator, we need that $\beta_1,\beta_2$ and $\beta_3$ are
non-negative and
\[
3 \beta_1 + 6 \beta_2 = 1, \quad 2 \beta_2 + 2 \beta_3 = 1.
\]
It is easy to see that the two equations above have solutions
bounded away from $0$ and $1$ and that the corresponding operator
has $r(T) < 1$. For example, choose $\beta_1=1/12$, $\beta_2 = 1/8$,
and $\beta_3 = 3/8$.
\end{proof}

\begin{lemma}\label{lemma:noise-alpha}
There exist a symmetric Markov operator $T$ on $\three^2$ such that
$r(T) < 1$ and such that if $T((x_1,x_2) \leftrightarrow (y_1,y_2)) > 0$ then
$x_1 \notin \{y_1,y_2\}$ and $y_1 \notin \{x_1,x_2\}$. Moreover, the noise operator $T$
satisfies the following property. Let $(x_1,x_2)$ be chosen according to the uniform distribution and $(y_1,y_2)$ be chosen
according $T$ applied to $(x_1,x_2)$. Then the distribution of $(x_2,y_2)$ is uniform.
\end{lemma}
\begin{proof}
The proof resembles the previous proof.
Again there are $3$ types of transitions.
\begin{itemize}
\item
With probability $\beta_1$ we have $(x,x) \leftrightarrow (y,y)$ where $x \neq
y$.
\item
With probability $\beta_2$ we have $(x,x) \leftrightarrow (y,z)$ where $x,y,z$ are
all different.
\item
With probability $\beta_3$ we have $(x,y) \leftrightarrow (z,y)$ where $x,y,z$ are all different.
\end{itemize}
For $T$ to be a symmetric Markov operator we require $\beta_1,\beta_2$ and $\beta_3$ to be
non-negative and
\[
2 \beta_1 + 2 \beta_2 = 1, \quad \beta_2 + \beta_3 = 1.
\]
Moreover, the last requirement of uniformity of $(x_2,y_2)$ amounts to the equation
\[
\beta_1/3 + 2\beta_2/3 = 2 \beta_3/3.
\]
It is easy to see that $\beta_2=\beta_3=0.5$ and $\beta_1=0$ is the
solution of all equations and that the corresponding operator has
$r(T) < 1$. This operator is illustrated in Figure \ref{fig:noiseoperators}(b).
\end{proof}

\subsection{The three reductions}\label{subsec:constr}

The basic idea in all three reductions is to take a label-cover instance
and to replace each vertex with a block of $q^R$ vertices, corresponding
to the $q$-ary hypercube $[q]^R$.
The intended way to $q$-color this block is by coloring $x \in [q]^R$ according to
$x_i$ where $i$ is the label given to this block. One can think of this coloring as an encoding of
the label $i$. We will essentially prove that any other
coloring of this block that uses relatively few colors, can
be ``list-decoded'' into at most $t$ labels from
$\sett{1,\ldots,R}$. By properly defining edges
connecting these blocks, we can guarantee that the lists
decoded from two blocks can be used as $t$-labelings for
the label-cover instance.

In the rest of this section, we use the following notation.
For a vector $x=(x_1,\ldots,x_n)$ and a permutation $\pi$ on $\{1,\ldots,n\}$,
we define $x^\pi = (x_{\pi(1)},\ldots,x_{\pi(n)})$.

\paragraph{$\a3c$:} Let $G=((V,E),R,\Psi)$ be a label-cover instance as in
Conjecture~\ref{conj:one-to-one}.
For $v \in V$ write $[v]$ for a collection of vertices, one per
point in $\three^{R}$. Let $e=(v,w)\in E$, and let $\psi$ be the
$\onetoone$-constraint associated with $e$. By
Definition~\ref{def:1-to-1} there is a permutation $\pi$ such that
$(a,b)\in \psi$ iff $b=\pi(a)$. We now write $[v,w]$ for the
following collection of edges. We put an edge $(x,y)$ for
$x=(x_1,\ldots,x_{R})\in [v]$ and $y=(y_1,\ldots,y_{R})\in [w]$ iff
$$\forall i\in \sett{1,\ldots,R},\quad T\left(x_{i}
\leftrightarrow y_{\pi (i)} \right)\neq 0$$
where $T$ is the noise
operator from Lemma~\ref{lemma:noise-almost}. In other words, $x$ is
adjacent to $y$ whenever
\[
T^{\otimes R}\left( x \leftrightarrow y^\pi \right) =
  \prod_{i=1}^R T\left(x_{i}
\leftrightarrow y_{\pi (i)} \right) \neq 0.\] The reduction
outputs the graph $[G] = ([V],[E])$ where $[V]$ is the disjoint
union of all blocks $[v]$ and $[E]$ is the disjoint union of
collection of the edges $[v,w]$.

\paragraph{$\col 4 Q$:} This reduction is nearly
identical to the one above, with the following changes:
\begin{itemize}
\item The starting point of the reduction is an instance $G = ((V,E),2R,\Psi)$ as in Conjecture~\ref{conj:two-to-two}.
\item Each vertex $v$ is replaced by a copy of $\four^{2R}$ (which we still denote $[v]$).
\item For every $(v,w)\in E$, let $\psi$ be the
$\twototwo$-constraint associated with $e$. By
Definition~\ref{def:2-to-2} there are two permutations $\pi_1,\pi_2$
such that $(a,b)\in \psi$ iff $(\pi_1^{-1}(a), \pi_2^{-1}(b))\in
\twototwo$. We now write $[v,w]$ for the following collection of
edges. We put an edge $(x,y)$ for $x=(x_1,\ldots,x_{2R})\in [v]$ and
$y=(y_1,\ldots,y_{2R})\in [w]$ if
$$\forall i\in \sett{1,\ldots,R},\quad T(( x_{\pi_1(2i-1)},x_{\pi_1(2i)}) \leftrightarrow
(y_{\pi_2(2i-1)},y_{\pi_2(2i)}))\neq 0$$ where $T$ is the noise
operator from Lemma~\ref{lemma:noise-2by2}. Equivalently, we put an edge if
$T^{\otimes R}(\bunch{x^{\pi_1}} \leftrightarrow \bunch{y^{\pi_2}}) \neq 0$.
\end{itemize}
As before, the reduction outputs the graph $[G] = ([V],[E])$ where
$[V]$ is the union of all blocks $[v]$ and $[E]$ is the union of
collection of the edges $[v,w]$.

\paragraph{$\col 3 Q$:}
Here again the reduction is nearly identical to the above, with the
following changes:
\begin{itemize}
\item The starting point of the reduction is an instance of label-cover, as in
Conjecture~\ref{conj:alpha}.
\item Each vertex $v$ is replaced by a copy of $\three^{2R}$ (which we again denote $[v]$).
\item For every $(v,w)\in E$, let $\pi_1,\pi_2$ be the permutations associated with the constraint, as in
Definition~\ref{def:alpha}. Define a collection $[v,w]$ of edges, by
including the edge $(x,y)\in [v]\times [w]$ iff
$$\forall i\in \sett{1,\ldots,R},\quad T(( x_{\pi_1(2i-1)},x_{\pi_1(2i)}) \leftrightarrow
(y_{\pi_2(2i-1)},y_{\pi_2(2i)}))\neq 0$$ where $T$ is the noise
operator from Lemma~\ref{lemma:noise-alpha}. As before, this condition can be written
as $T^{\otimes R}(\bunch{x^{\pi_1}} \leftrightarrow \bunch{y^{\pi_2}}) \neq 0$.
\end{itemize}
As before, we look at the coloring problem of the graph $[G] =
([V],[E])$ where $[V]$ is the union of all blocks $[v]$ and $[E]$ is
the union of collection of the edges $[v,w]$.

\subsection{Completeness of the three reductions}\label{subsec:complete}

\paragraph{$\a3c$:}
If $\isat(G)\ge 1-\eps$, then there is some $S\subseteq V$ of size
$(1-\eps)\card V$ and a labeling $\ell:S\to R$ that satisfies all of
the constraints induced by $S$.
We $3$-color all of the vertices in $\cup_{v\in S}[v]$ as follows.
Let $c:\cup_{v\in S}[v] \to \three$ be defined as follows. For every
$v\in S$, the color of $x = (x_1,\ldots,x_R) \in \three^R=[v]$ is
defined to be $c(x) \defeq x_i$, where $i = \ell(v)\in
\sett{1,\ldots,R}$.

To see that $c$ is a legal coloring on
$\cup_{v\in S}[v]$, observe that if $x\in [v]$ and $y\in[w]$ share
the same color, then $x_{i}=y_j$ for $i=\ell(v)$ and $j=\ell(w)$.
Since $\ell$ satisfies every constraint induced by $S$, it follows that if
$(v,w)$ is a constraint with an associated permutation $\pi$, then $j = \pi(i)$. Since
$T(z \leftrightarrow z)=0$ for all $z\in\three$, there is no edge
between $x$ and $y$.

\paragraph{$\col 4 Q$:}
Let $\ell:V\to \sett{1,\ldots,2R}$ be a labeling that satisfies all
the constraints in $G$. We define a legal $4$-coloring
$c:[V]\to\four$ as follows. For a vertex $x = (x_1,\ldots,x_{2R})
\in \four^{2R}=[v]$ set $c(x)\defeq x_i$, where $i = \ell(v)\in
\sett{1,\ldots,2R}$.

To see that $c$ is a legal coloring, fix any
$\twototwo$ constraint $(v,w)\in E$ and let $\pi_1,\pi_2$ be the permutations
associated with it. Let $i=\ell(v)$ and
$j=\ell(w)$, so by assumption $(\pi_1^{-1}(i),\pi_2^{-1}(j))\in
\twototwo$. In other words there is some $k\in \sett{1,\ldots,R}$
such that $i\in\sett{\pi_1(2k-1),\pi_1(2k)}$ and
$j\in\sett{\pi_2(2k-1),\pi_2(2k)}$.
If $x\in [v]$
and $y\in[w]$ share the same color, then $x_i = c(x) = c(y) = y_j$.
Since
$$x_{i}\in\sett{x^{\pi_1}_{2k-1},x^{\pi_1}_{2k}}\quad
   \hbox{and}\quad
y_{j}\in\sett{y^{\pi_2}_{2k-1},y^{\pi_2}_{2k}}$$
we have  that the above sets intersect. This, by
Lemma~\ref{lemma:noise-2by2}, implies that
$T^{\otimes R}(\bunch{x^{\pi_1}} \leftrightarrow \bunch{y^{\pi_2}})=0$. So the vertices $ x, y$
cannot be adjacent, hence the coloring is legal.

\paragraph{$\col 3 Q$:}
Here the argument is nearly identical to the above. Let $\ell:V\to
\sett{1,\ldots,2R}$ be a labeling that satisfies all of the
constraints in $G$. We define a legal $3$-coloring $c:[V]\to\three$
like before: $c( x)\defeq x_i$, where $i = \ell(v)\in
\sett{1,\ldots,2R}$. To see that $c$ is a legal coloring, fix any
edge $(v,w)\in E$ and let $\pi_1,\pi_2$ be the permutations
associated with the $\fish$-constraint. Let $i=\ell(v)$ and
$j=\ell(w)$, so by assumption $(\pi^{-1}_1(i),\pi^{-1}_2(j))\in
\fish$. In other words there is some $k\in \sett{1,\ldots,R}$ such
that $i\in\sett{\pi_1(2k-1),\pi_1(2k)}$ and
$j\in\sett{\pi_2(2k-1),\pi_2(2k)}$ and not both $i=\pi_1(2k)$ and
$j=\pi_2(2k)$. Assume, without loss of generality, that $i=\pi_1(2k-1)$, so
$x_{i}=x^{\pi_1}_{2k-1}$ and $y_{j} \in\sett{ y^{\pi_2}_{2k-1},
y^{\pi_2}_{2k}}$.

If $x\in [v]$ and $y\in[w]$ share the same color, then
$x_i = c(x) = c(y) = y_j$, so
$$x^{\pi_1}_{2k-1} = x_{i}=y_{j} \in
\sett{y^{\pi_2}_{2k-1},y^{\pi_2}_{2k}}\,.$$ By
Lemma~\ref{lemma:noise-alpha} this implies $T((
x^{\pi_1}_{2k-1},x^{\pi_1}_{2k})\leftrightarrow(y^{\pi_2}_{2k-1},y^{\pi_2}_{2k}))=0$,
which means there is no edge between $x$ and $y$.

\subsection{Soundness of the three reductions}\label{subsec:sound}
\def\IS{{S}}

Before presenting the soundness proofs, we need the following corollary.
It is simply a special case of Theorem~\ref{thm:fourier}
stated in the contrapositive, with $\eps$ playing the role of $\nu$ and $\mu$.
Here we use the fact that $\ip{F_\eps,U_\rho(1-F_{1-\eps})}_\gamma > 0$
whenever $\eps>0$.
\begin{corollary}\label{cor:main}
Let $q$ be a fixed integer and let $T$ be a reversible Markov
operator on $[q]$ such that $r(T) < 1$.
For every $\eps>0$ there exist $\delta>0$ and $k\in\mathbb{N}$ such
that the following holds.
For any $f,g:[q]^n\to [0,1]$, if $E[f]>\eps$, $E[g]>\eps$, and $\ip{f,Tg}=0$, then
$$\exists i\in \{1,\ldots,n\},\quad I^{\le k}_i(f) \ge \delta\quad\hbox{ and }\quad
I^{\le k}_i(g) \ge \delta\,.$$
\end{corollary}

\paragraph{$\a3c$:}
We will show that if $[G]$ has an independent set $S\subseteq [V]$
of relative size $\ge 2\eps$, then $\isat_t(G)\ge \eps$ for a fixed
constant $t>0$ that depends only on $\eps$. More explicitly, we will
find a set $J\subseteq V$, and a $t$-labeling $L:J\to {R \choose
{\le t}}$ such that $\card J \ge \eps \card V$ and $L$ satisfies all
the constraints of $G$ induced by $J$. In other words, for every
constraint $\c$ over an edge $(u,v)\in E\cap J^2$, there are values
$a\in L(u)$ and $b\in L(v)$ such that $(a,b)\in \c$.

Let $J$ be the set of all vertices $v \in V$ such that the fraction
of vertices belonging to $\IS$ in $[v]$ is at least $\eps$. Then,
since $\card S \ge 2\eps\card{[V]}$, Markov's inequality implies
$\card J\ge\eps \card V$.

For each $v\in J$ let $f_v:\three^R\to\sett{0,1}$ be the
characteristic function of S restricted to $[v]$, so $\E{[f_v]}\ge
\eps$. Select $\delta,k$ according to Corollary~\ref{cor:main} with $\eps$ and
the operator $T$ of Lemma~\ref{lemma:noise-almost}, and
set
$$L(v) = \set{i \in \sett{1,\ldots,R}}{I_i^{\le k}(f_v) \ge \delta}\,.$$ Clearly,
$\card{L(v)} \le k/\delta$ because $\sum_{i=1}^{R} I^{\le k}_i(f)
\le k$. Thus, $L$ is a $t$-labeling for $t= k/ \delta$. The main
point to prove is that for every edge $e=(v_1,v_2)\in E\cap J^2$
induced on $J$, there is some $a\in L(v_1)$ and $b\in L(v_2)$ such
that $(a,b)\in \psi_e$. In other words, $\isat_t(G)\ge \card J/\card
V \ge \eps$.

Fix $(v_1,v_2)\in E\cap J^2$, and let $\pi$ be the permutation
associated with the $\onetoone$ constraint on this edge. (It may be
easier to first think of $\pi=id$.) Recall that the edges in
$[v_1,v_2]$ were defined based on $\pi$, and on the noise operator
$T$ defined in Lemma~\ref{lemma:noise-almost}.
Let $f = f_{v_1}$, and define $g$
by $g(x^\pi) = f_{v_2}(x)$. Since $\IS$ is an independent set,
$f(x)=f_{v_1}(x) = 1$ and $g(y^\pi )=f_{v_2}(y)=1$ implies that
$x,y$ are not adjacent, so by construction $T(x\leftrightarrow
y^\pi)=0$. Therefore,

$$\ip{f,Tg} = {3^{-R}}\sum_x f(x)Tg(x) = {3^{-R}} \sum_x f(x)\sum_{y^\pi} T(x
\leftrightarrow y^\pi)g(y^\pi) = \sum_{x,y^\pi}0 = 0\,.$$ Also, by
assumption, $E[g]\ge \eps$ and $E [f] \ge \eps $.
Corollary~\ref{cor:main} implies that there is some index $i \in
\sett{1,\ldots,R}$ for which both $I^{\le k}_i(f) \ge \delta$ and
$I^{\le k}_i(g)
\ge \delta$. By definition of $L$, $i\in L(v_1)$. Since the $i$-th variable in $g$ is the $\pi(i)$-th variable in
$f_{v_2}$, $\pi(i)\in L(v_2)$. It follows that there are values
$i\in L(v_1)$ and $\pi(i)\in L(v_2)$ such that $(i,\pi(i))$
satisfies the constraint on $(v_1,v_2)$. This means that $\isat_t(G)
\ge \card J / \card V \ge \eps$.

\paragraph{$\col 4 Q$:}
We outline the argument and emphasize only the modifications. Assume
that $[G]$ contains an
independent set $S\subseteq[V]$ whose relative size is at least
$1/Q$ and set $\eps = 1/2Q$.
\begin{itemize}
\item Let $f_v:\four^{2R}\to \sett{0,1}$ be the characteristic function of $S$ in
$[v]$. Define the set $J\subseteq V$ as before and for all $v\in J$, define
$$L(v) = \set{i \in \sett{1,\ldots,2R}}{I_i^{\le 2k}(f_v) \ge \frac\delta 2
}$$
where $k,\delta$ are the values given by Corollary \ref{cor:main} with $\eps$ and
the operator $T$ of Lemma~\ref{lemma:noise-2by2}.
As before, $\card J\ge \eps\card V$ and $\E{[f_v]}\ge\eps$ for
$v\in J$. Now $L$ is a $t$-labeling with $t = 4k/\delta$. Fix an edge $(v,w)\in E\cap J^2$ and
let $\pi_1,\pi_2$ be the associated permutations. Define $f,g$ by
$f(x^{\pi_1})
\defeq  f_{v_1}(x)$ and $g(y^{\pi_2})
\defeq f_{v_2}(y)$.
\item Since $\IS$ is an independent set,
$f(x^{\pi_1})=f_{v_1}(x) = 1$ and $g(y^{\pi_2})=f_{v_2}(y)=1$
implies that $x,y$ are not adjacent, so by construction
$T(x^{\pi_1}\leftrightarrow y^{\pi_2})=0$. Therefore, $\ip{f,Tg}=0$.

\item Now, recalling Definition~\ref{def:bunch}, consider the
functions $\bunch f,\bunch g:(\four^2)^R\to\sett{0,1}$. Applying
Corollary~\ref{cor:main} on $\bunch f,\bunch g$ we may deduce the
existence of an index $i \in \sett{1,\ldots,R}$ for which both
$I^{\le k}_i(\bunch f) \ge \delta$ and $I^{\le k}_i(\bunch g) \ge \delta$.
By Claim~\ref{claim:two-vars}, $\delta \le I^{\le k}_i(\bunch f)\le
I^{\le 2k}_{2i-1}(f)+ I^{\le 2k}_{2i}(f)$, so either $I^{\le
  2k}_{2i-1}(f) \geq \delta/2$ or $I^{\le 2k}_{2i}(f) \geq \delta/2$.
Since the $j$-th
variable in $f$ is the $\pi_1(j)$-th variable in $f_{v_1}$, this
puts either $\pi_1(2i)$ or $\pi_1(2i-1)$ in $L(v_1)$. Similarly, at
least one of $\pi_2(2i),\pi_2(2i-1)$ is in $L(v_2)$. Thus, there are
$a\in L(v_1)$ and $b\in L(v_2)$ such that
$(\pi^{-1}_1(a),\pi^{-1}_2(b))\in \twototwo$ so $L$ satisfies the
constraint on $(v_1,v_2)$.
\end{itemize}
We have shown that $L$ satisfies every constraint induced by $J$, so
$\isat_t(G) \ge \eps$.

\paragraph{$\col 3 Q$:}
The argument here is similar to the previous one.
The main difference is in the third step, where we replace
Corollary~\ref{cor:main} by the following corollary of
Theorem~\ref{thm:fish_fourier}. The corollary follows by letting $\eps$ play the role of $\mu$ and $\nu$,
and using the fact that $\ip{F_\eps,U_\rho(1-F_{1-\eps})}_\gamma
> 0$ whenever $\eps>0$.
\begin{corollary}\label{cor:main-fish}
Let $T$ be the operator on $\three^2$ defined in
Lemma~\ref{lemma:noise-alpha}. For any $\eps
> 0$, there exists $\delta > 0,k\in\mathbb{N}$, such that for
any functions $f,g : \three^{2R} \to [0,1]$ satisfying $\E[f] \ge
\eps, \E[g] \ge\eps$, there exists some $i\in \sett{1,\ldots,R}$
such that either
\begin{equation*}
\min \big( I_{2i-1}^{\leq k}(f),I_{2i-1}^{\leq k}(g) \big) \ge \delta \quad
\text{or}
\quad \min \big( I_{2i-1}^{\leq k}(f),I_{2i}^{\leq k}(g) \big) \ge \delta \quad
\text{or}
\quad \min \big( I_{2i}^{\leq k}(f),I_{2i-1}^{\leq k}(g) \big) \ge \delta.
\end{equation*}
\end{corollary}

Now we have functions $f_{v}:\three^{2R}\to\sett{0,1}$, and $J$ is defined
as before. Define a labeling
$$L(v) = \set{i \in \sett{1,\ldots,2R}}{I_i^{\le k}(f_v) \ge \delta
}$$
where $k,\delta$ are the values given by Corollary \ref{cor:main-fish} with $\eps$.
Then $L$ is a $t$-labeling with $t=k/\delta$.

Let us now show that $L$ is a satisfying $t$-labeling. Let $(v_1,v_2)$ be
a $\fish$-constraint with associated permutations $\pi_1,\pi_2$. Define
$f(x^{\pi_1})=f_{v_1}(x), g(x^{\pi_2})=f_{v_2}(x)$. We apply
Corollary~\ref{cor:main-fish} on $f,g$, and obtain
an index $i \in \{1,\ldots,R\}$. Since the $j$-th variable in $f$ is
the $\pi_1(j)$-th variable in $f_{v_1}$, this puts either
$\pi_1(2i)$ or $\pi_1(2i-1)$ in $L(v_1)$. Similarly, at least one of
$\pi_2(2i),\pi_2(2i-1)$ is in $L(v_2)$. Moreover, we are guaranteed
that either $\pi_1(2i-1)\in L(v_1)$ or $\pi_2(2i-1)\in L(v_2)$.
Thus, there are $a\in L(v_1)$ and $b\in L(v_2)$ such that
$(\pi^{-1}_1(a),\pi^{-1}_2(b))\in \fish$ so $L$ satisfies the
constraint on $(v_1,v_2)$.

\subsection*{Acknowledgements}

We thank Uri Zwick for his help with the literature.

\bibliography{coloring}
\bibliographystyle{abbrv}

\appendix

\section{Comparison with Khot's Conjectures}\label{app:Conjectures}

Let us
first state Khot's original conjectures. For $d \ge 1$, an instance
of the weighted bipartite $d$-to-$1$ label cover problem is given by
a tuple $\Phi=(X,Y,\Psi,W)$. We often refer to vertices in $X$ as
{\em left} vertices and to vertices in $Y$ as {\em right} vertices.
The set $\Psi$ consists of one $d$-to-$1$ relation $\psi_{xy}$ for
each $x\in X$ and $y\in Y$. More precisely, $\psi_{xy} \subseteq
\{1,\ldots,R\} \times \{1,\ldots,R/d\}$ is such that for any $b\in \{1,\ldots,R/d\}$ there are precisely
$d$ elements $a\in \{1,\ldots,R\}$ such that $(a,b) \in \psi_{xy}$. The set
$W$ includes a non-negative weight $w_{xy}\ge 0$ for each $x\in X$,
$y\in Y$. We denote by $w(\Phi, x)$ the sum $\sum_{y\in Y} w_{xy}$
and by $w(\Phi)$ the sum $\sum_{x\in X,y\in Y} w_{xy}$. A {\em
labeling} is a function $L$ mapping $X$ to $\{1,\ldots,R\}$ and $Y$ to $\{1,\ldots,R/d\}$.
A constraint $\psi_{xy}$ is {\em satisfied} by a labeling $L$ if
$(L(x),L(y)) \in \psi_{xy}$. Also, for a labeling $L$, the weight of
satisfied constraints, denoted by $w_L(\Phi)$, is $\sum w_{xy}$
where the sum is taken over all $x\in X$ and $y\in Y$ such that
$\psi_{xy}$ is satisfied by $L$. Similarly, we define $w_L(\Phi, x)$
as $\sum w_{xy}$ where the sum is now taken over all $y\in Y$ such
that $\psi_{xy}$ is satisfied by $L$. The following conjectures were
presented in~\cite{Khot-unique-games}.

\begin{conjecture}[Bipartite $1$-to-$1$ Conjecture]\label{unique_games_conjecture}
For any $\zeta, \gamma>0$ there exists a constant $R$ such that the
following is NP-hard. Given a $1$-to-$1$ label cover instance $\Phi$
with label set $\{1,\ldots,R\}$ and $w(\Phi)=1$ distinguish between the case
where there exists a labeling $L$ such that $w_L(\Phi) \ge
1-\zeta$ and the case where for any labeling $L$, $w_L(\Phi) \le
\gamma$.
\end{conjecture}

In the following conjecture, $d$ is any fixed integer greater than
$1$.

\begin{conjecture}[Bipartite $d$-to-$1$ Conjecture]\label{dto1_conjecture}
For any $\gamma>0$ there exists a constant $R$ such that the
following is NP-hard. Given a bipartite $d$-to-$1$ label cover
instance $\Phi$ with label sets $\{1,\ldots,R\},\{1,\ldots,R/d\}$ and $w(\Phi)=1$
distinguish between the case where there exists a labeling $L$
such that $w_L(\Phi) = 1$ and the case where for any labeling $L$,
$w_L(\Phi) \le \gamma$.
\end{conjecture}

The theorem we prove in this section is the following.

\begin{theorem}\label{theorem_PCP}
Conjecture \ref{conj:one-to-one} follows from Conjecture
\ref{unique_games_conjecture} and Conjecture
\ref{conj:two-to-two} follows from Conjecture
\ref{dto1_conjecture} for $d=2$.\footnote{We in fact show that
for any $d \ge 2$, the natural extension of Conjecture \ref{conj:two-to-two}
to $d$-to-$d$ constraints follows from Conjecture \ref{dto1_conjecture} with
the same value of $d$.}
\end{theorem}

The proof follows by combining Lemmas \ref{lemma_first},
\ref{lemma_second}, \ref{lemma_third}, and \ref{lemma_fourth}. Each
lemma presents an elementary transformation between variants of the
label cover problem.
The first transformation modifies a bipartite
label cover instance so that all $X$ variables have the same weight.
When we say below that $\Phi'$ has the same type of constraints as
$\Phi$ we mean that the transformation only duplicates existing
constraints and hence if $\Phi$ consists of $d$-to-$1$ constraints
for some $d\ge 1$, then so does $\Phi'$.

\begin{lemma}\label{lemma_first}
There exists an efficient procedure that given a weighted bipartite
label cover instance $\Phi=(X,Y,\Psi,W)$ with $w(\Phi)=1$ and a
constant $\ell$, outputs a weighted bipartite label cover instance
$\Phi'=(X',Y,\Psi',W')$ on the same label sets and with the same
type of constraints with the following properties:
\begin{itemize}
 \item
 For all $x\in X'$, $w(\Phi',x)=1$.
 \item
 For any $\zeta\ge 0$, if there exists a labeling $L$ to $\Phi$ such that $w_L(\Phi) \ge 1-\zeta$
 then there exists a labeling $L'$ to $\Phi'$ in which $1-\sqrt{(1 + \frac{1}{\ell-1})\zeta}$ of the variables $x$ in $X'$
 satisfy that $w_{L'}(\Phi',x) \ge 1-\sqrt{(1 + \frac{1}{\ell-1})\zeta}$. In particular, if there exists a labeling $L$ such
 that $w_L(\Phi) = 1$
 then there exists a labeling $L'$ in which all variables satisfy $w_{L'}(\Phi',x) = 1$.
 \item
 For any $\beta_2, \gamma > 0$, if there exists a labeling $L'$ to $\Phi'$ in which
 $\beta_2$ of the variables $x$ in $X'$ satisfy $w_{L'}(\Phi',x) \ge \gamma$, then there
 exists a labeling $L$ to $\Phi$ such that $w_L(\Phi) \ge (1-\frac{1}{\ell}) \beta_2 \gamma$.
\end{itemize}
\end{lemma}
\begin{proof}
Given $\Phi$ as above, we define $\Phi'=(X',Y,\Psi',W')$ as follows.
The set $X'$ includes $k(x)$ copies of each $x\in X$,
$x^{(1)},\ldots,x^{(k(x))}$ where $k(x)$ is defined as $\lfloor
\ell\cdot |X| \cdot w(\Phi, x) \rfloor$. For every $x\in X$, $y\in Y$
and $i\in \{1,\ldots,k(x)\}$ we define $\psi'_{x^{(i)}y}$ as
$\psi_{xy}$ and the weight $w'_{x^{(i)}y}$ as $w_{x y} / w(\Phi,
x)$. Notice that $w(\Phi', x)=1$ for all $x\in X'$ and that
$(\ell-1)|X| \le |X'| \le \ell|X|$. Moreover, for any $x\in X$, $y\in Y$,
the total weight of constraints created from $\psi_{xy}$ is $k(x)
w_{xy} / w(\Phi,x) \le \ell |X| w_{xy}$.

We now prove the second property. Given a labeling $L$ to $\Phi$
that satisfies constraints of weight at least $1-\zeta$, consider
the labeling $L'$ defined by $L'(x^{(i)})=L(x)$ and $L'(y)=L(y)$.
By the property mentioned above, the total weight of unsatisfied
constraints in $\Phi'$ is at most $\ell|X|\zeta$. Since the total
weight in $\Phi'$ is at least $(\ell-1)|X|$, we obtain that the
fraction of unsatisfied constraints is at most
$(1+\frac{1}{\ell-1})\zeta$. Hence, by a Markov argument, we obtain
that for at least $1-\sqrt{(1 + \frac{1}{\ell-1})\zeta}$ of the $X'$
variables $w_{L'}(\Phi',x) \ge 1-\sqrt{(1 + \frac{1}{\ell-1})\zeta}.$

We now prove the third property. Assume we are given a labeling
$L'$ to $\Phi'$ for which $\beta_2$ of the variables have
$w_{L'}(\Phi',x) \ge \gamma$. Without loss of generality we can
assume that for every $x\in X$, the labeling $L'(x^{(i)})$ is the
same for all $i$. This holds since the constraints between $x^{(i)}$
and the $Y$ variables are the same for all $i\in \{1,\ldots,k(x)\}$.
We define the labeling $L$ as $L(x)=L'(x^{(1)})$. The weight of
constraints satisfied by $L$ is:
\begin{align*}
 \sum_{x\in X} w_{L}(\Phi,x) &\ge \frac{1}{\ell |X|} \sum_{x\in X} k(x) \cdot w_{L}(\Phi,x) / w(\Phi,x) \\
 & = \frac{1}{\ell |X|} \sum_{x\in X'} w_{L'}(\Phi',x) \\
 & \ge \frac{1}{\ell |X|} \beta_2 |X'| \gamma \ge
    \Big(1 - \frac{1}{\ell}\Big) \beta_2 \gamma
\end{align*}
where the first inequality follows from the definition of $k(x)$.
\end{proof}

The second transformation creates an {\em unweighted} label cover
instance. Such an instance is given by a tuple $\Phi=(X,Y,\Psi,E)$.
The multiset $E$ includes pairs $(x,y) \in X \times Y$ and we can
think of $(X,Y,E)$ as a bipartite graph (possibly with parallel
edges). For each $e\in E$, $\Psi$ includes a constraint, as before.
The instances created by this transformation are left-regular, in
the sense that the number of constraints $(x,y)\in E$ incident to
each $x \in X$ is the same.

\begin{lemma}\label{lemma_second}
There exists an efficient procedure that given a weighted bipartite
label cover instance $\Phi=(X,Y,\Psi,W)$ with $w(\Phi,x)=1$ for all
$x\in X$ and a constant $\ell$, outputs an unweighted bipartite label
cover instance $\Phi'=(X,Y,\Psi',E')$ on the same label sets and
with the same type of constraints with the following properties:
\begin{itemize}
 \item
 All left degrees are equal to $\alpha = \ell|Y|$.
 \item
 For any $\beta,\zeta > 0$, if there exists a labeling $L$ to $\Phi$ such that $w_L(\Phi,x) \ge 1-\zeta$
 for at least $1-\beta$ of the variables in $X$, then there exists a labeling $L'$ to $\Phi'$
 in which for at least $1-\beta$ of the variables in $X$, at least $1-\zeta-1/\ell$ of their
 incident constraints are satisfied.
 Moreover, if there exists a labeling $L$ such that $w_L(\Phi,x) = 1$
 for all $x$ then there exists a labeling $L'$ to $\Phi'$ that satisfies all constraints.
 \item
 For any $\beta, \gamma > 0$, if there exists a labeling $L'$ to $\Phi'$ in which
 $\beta$ of the variables in $X$ have $\gamma$ of their incident constraints satisfied, then there
 exists a labeling $L$ to $\Phi$ such that for $\beta$ of the variables in $X$, $w_L(\Phi,x) > \gamma - 1/\ell$.
\end{itemize}
\end{lemma}
\begin{proof}
We define the instance $\Phi'=(X,Y,\Psi',E')$ as follows. For each
$x\in X$, choose some $y_0(x) \in Y$ such that $w_{xy_0(x)}>0$. For
every $x\in X$, $y\neq y_0(x)$, $E'$ contains $\lfloor \alpha w_{xy}
\rfloor$ edges from $x$ to $y$ associated with the constraint
$\psi_{xy}$. Moreover, for every $x\in X$, $E'$ contains $\alpha -
\sum_{y\in Y \setminus \{y_0(x)\}} \lfloor \alpha w_{xy} \rfloor$
edges from $x$ to $y_0(x)$ associated with the constraints
$\psi_{xy_0(x)}$. Notice that all left degrees are equal to
$\alpha$. Moreover, for any $x$, $y\neq y_0(x)$, we have that the
number of edges between $x$ and $y$ is at most $\alpha w_{xy}$ and
the number of edges from $x$ to $y_0(x)$ is at most $\alpha w_{x
y_0(x)} + |Y| = \alpha (w_{x y_0(x)} + 1/\ell)$.

Consider a labeling $L$ to $\Phi$ and let $x\in X$ be such that
$w_L(\Phi, x) > 1-\zeta$. Then, in $\Phi'$, the same labeling
satisfies that the number of incident constraints to $x$ that are
satisfied is at least $(1-\zeta - 1/\ell)\alpha$. Moreover, if
$w_L(\Phi,x)=1$ then all its incident constraints in $\Phi'$ are
satisfied (this uses that $w_{xy_0(x)}>0$). Finally, consider a labeling $L'$ to $\Phi'$ and let $x\in X$ have $\gamma$ of their
incident constraints satisfied. Then, $w_{L'}(\Phi,x) > \gamma -
\frac{1}{\ell}$.
\end{proof}

In the third lemma we modify a left-regular unweighted label cover
instance so that it has the following property: if there exists a labeling to the original instance that for many variables
satisfies many of their incident constraints, then the resulting
instance has a labeling that for many variables satisfies {\em
all} their incident constraints. But first, we prove a combinatorial
claim.

\begin{claim}\label{clm:pairwisedisjoint}
For any integers $\ell,d,R$ and real $0 < \gamma < \frac{1}{\ell^2 d}$,
let $\calF \subseteq P(\{1,\ldots,R\})$ be a multiset containing subsets of
$\{1,\ldots,R\}$ each of size at most $d$ with the property that no element
$i\in \{1,\ldots,R\}$ is contained in more than $\gamma$ fraction of the sets
in $\calF$. Then, the probability that a sequence of sets
$F_1,F_2,\ldots,F_\ell$ chosen uniformly from $\calF$ (with
repetitions) is pairwise disjoint is at least $1-\ell^2 d\gamma$.
\end{claim}
\begin{proof}
Note that by the union bound it suffices to prove that
$\Pr[F_1 \cap F_2
\neq \emptyset] \leq d \gamma$. This follows by fixing $F_1$ and using the
union bound again:
$$\Pr[F_1 \cap F_2 \neq \emptyset] \leq \sum_{x \in F_1} \Pr[x \in F_2] \leq d \gamma.$$
\end{proof}

\begin{lemma}\label{lemma_third}
There exists an efficient procedure that given an unweighted
bipartite $d$-to-$1$ label cover instance $\Phi=(X,Y,\Psi,E)$ with
all left-degrees equal to some $\alpha$, and a constant $\ell$, outputs
an unweighted bipartite $d$-to-$1$ label cover instance
$\Phi'=(X',Y,\Psi',E')$ on the same label sets with the following
properties:
\begin{itemize}
 \item
 All left degrees are equal to $\ell$.
 \item
 For any $\beta,\zeta \ge 0$, if there exists a labeling $L$ to $\Phi$ such that
 for at least $1-\beta$ of the variables in $X$ $1-\zeta$ of their incident constraints
 are satisfied, then there exists a labeling $L'$ to $\Phi'$
 in which $(1-\zeta)^\ell (1-\beta)$ of the $X'$ variables have all their $\ell$ constraints
 satisfied.
 In particular, if there exists a labeling $L$ to $\Phi$ that satisfies all constraints
 then there exists a labeling $L'$ to $\Phi'$ that satisfies all constraints.
 \item
 For any $\beta>0$, $0<\gamma < \frac{1}{\ell^2 d}$, if in any labeling $L$ to $\Phi$ at most
 $\beta$ of the variables have $\gamma$ of their incident constraints satisfied, then in any labeling
 $L'$ to $\Phi'$, the fraction of satisfied constraints is at most
 $\beta  + \frac{1}{\ell} + (1-\beta) \ell^2d\gamma$.
\end{itemize}
\end{lemma}
\begin{proof}
We define $\Phi'=(X',Y,\Psi',E')$ as follows. For each $x\in X$,
consider its neighbors $(y_1,\ldots,y_\alpha)$ listed with
multiplicities. For each sequence $(y_{i_1},\ldots,y_{i_\ell})$ where
$i_1,\ldots,i_\ell \in \{1,\ldots,\alpha\}$ we create a variable in
$X'$. This variable is connected to $y_{i_1},\ldots,y_{i_\ell}$ with
the same constraints as $x$, namely
$\psi_{xy_{i_1}},\ldots,\psi_{xy_{i_\ell}}$. Notice that the total
number of variables created from each $x\in X$ is $\alpha^\ell$. Hence,
$|X'| = \alpha^\ell |X|$.

We now prove the second property. Assume that $L$ is a labeling
to $\Phi$ such that for at least $1-\beta$ of the variables in $X$,
$1-\zeta$ of their incident constraints are satisfied. Let $L'$ be
the labeling to $\Phi'$ assigning to each of the variables created
from $x\in X$ the value $L(x)$ and for each $y\in Y$ the value
$L(y)$. Consider a variable $x\in X$ that has $1-\zeta$ of its
incident constraints satisfied and let $Y_x$ denote the set of
variables $y\in Y$ such that $\psi_{xy}$ is satisfied. Then among
the variables in $X'$ created from $x$, the number of variables that
are connected only to variables in $Y_x$ is at least $\alpha^\ell
(1-\zeta)^\ell$.
 Therefore, the total number of variables all of whose constraints are satisfied by $L'$ is at least
 $$ \alpha^\ell (1-\zeta)^\ell (1-\beta) |X| = (1-\zeta)^\ell (1-\beta) |X'|.$$

We now prove the third property. Assume that in any labeling $L$
to $\Phi$ at most $\beta$ of the $X$ variables have $\gamma$ of
their incident constraints satisfied. Let $L'$ be an arbitrary
labeling to $\Phi'$. For each $x\in X$ define $\calF_x \subseteq
P(\{1,\ldots,R\})$ as the multiset that contains for each constraint incident
to $x$ the set of labels to $x$ that, together with the
labeling to the $Y$ variables given by $L'$, satisfy this
constraint. So $\calF_x$ contains $\alpha$ sets, each of size $d$.
Moreover, our assumption above implies that for at least $1-\beta$
of the variables $x\in X$, no element $i\in \{1,\ldots,R\}$ is contained in
more than $\gamma$ fraction of the sets in $\calF_x$. By Claim
\ref{clm:pairwisedisjoint}, for such $x$, at least $1-\ell^2 d\gamma$
fraction of the variables in $X'$ created from $x$ have the property
that it is impossible to satisfy more than one of their incident
constraints simultaneously. Hence, the number of constraints in
$\Phi'$ satisfied by $L'$ is at most
\begin{align*}
 & \alpha^\ell \cdot \beta \cdot |X| \cdot \ell + \alpha^\ell (1-\beta)|X|
  \Big(  (1-\ell^2 d\gamma) + (\ell^2d\gamma)\cdot \ell \Big)  \\
 &= |X'| \left( \beta \ell + (1-\beta) (1-\ell^2 d\gamma) + (1-\beta) (\ell^2d\gamma) \ell \right) \\
 & \le |E'| \left( \beta  + \frac{1}{\ell} + (1-\beta) \ell^2d\gamma \right).
\end{align*}
\end{proof}

The last lemma transforms a bipartite label cover into a
non-bipartite label cover. This transformation no longer preserves
the constraint type: $d$-to-$1$ constraints become $d$-to-$d$
constraints. We first prove a simple combinatorial claim. 

\begin{claim}\label{clm:pairwiseintersecting}
Let $A_1,\ldots,A_N$ be a sequence of pairwise intersecting sets of
size at most $T$. Then there exists an element contained in at least
$N/T$ of the sets.
\end{claim}
\begin{proof}
All sets intersect $A_1$ in at least one element. Since $|A_1|\le
T$, there exists an element of $A_1$ contained in at least $N/T$ of
the sets.
\end{proof}

For the following lemma, recall from Definition
\ref{def:multi-labeling} that a $t$-labeling labels each vertex with
a set of at most $t$ labels. Recall also that a constraint on $x,y$
is satisfied by a $t$-labeling $L$ if there is a label $a\in L(x)$
and $b\in L(y)$ such that $(a,b)$ satisfies the constraint.

\begin{lemma}\label{lemma_fourth}
There exists an efficient procedure that given an unweighted
bipartite $d$-to-$1$ label cover instance $\Phi=(X,Y,\Psi,E)$ on
label sets $\{1,\ldots,R\},\{1,\ldots,R/d\}$, with all left-degrees equal to some $\ell$,
outputs an unweighted $d$-to-$d$ label cover instance
$\Phi'=(X,\Psi',E')$ on label set $\{1,\ldots,R\}$ with the following
properties:
\begin{itemize}
 \item
 For any $\beta \ge 0$, if there exists a labeling $L$ to $\Phi$
 in which $1-\beta$ of the $X$ variables have all their $\ell$ incident constraints
 satisfied, then there exists a labeling to $\Phi'$ and a set of $1-\beta$
 of the variables of $X$ such that all the constraints between them are satisfied.
 In particular, if there exists a labeling $L$ to $\Phi$ that satisfies all constraints
 then there exists a labeling $L'$ to $\Phi'$ that satisfies all constraints.
 \item
 For any $\beta>0$ and integer $t$, if there exists a $t$-labeling $L'$ to $\Phi'$
 and a set of $\beta$ variables of $X$ such that all the constraints between them are satisfied,
 then there exists a labeling $L$ to $\Phi$ that satisfies at least $\beta/t^2$ of the constraints.
\end{itemize}
\end{lemma}
\begin{proof}
For each pair of constraints $(x_1,y),(x_2,y)\in E$ that share a $Y$
variable we add one constraint $(x_1,x_2)\in E'$. This constraint is
satisfied when there exists a labeling to $y$ that agrees with
the labeling to $x_1$ and $x_2$. More precisely,
$$ \psi'_{x_1x_2} = \Big\{ (a_1,a_2) \in \{1,\ldots,R\} \times \{1,\ldots,R\} ~\Big|~
     \exists b \in \{1,\ldots,R/d\}~(a_1,b) \in \psi_{x_1 y} \wedge (a_2,b) \in \psi_{x_2 y} \Big\} .$$
Notice that if the constraints in $\Psi$ are $d$-to-$1$ then the
constraints in $\Psi'$ are $d$-to-$d$.

We now prove the second property. Let $L$ be a labeling to $\Phi$
and let $C \subseteq X$ be of size $|C| \ge (1-\beta)|X|$ such that
all constraints incident to variables in $C$ are satisfied by $L$.
Consider the labeling $L'$ to $\Phi'$ given by $L'(x)=L(x)$. Then,
we claim that $L'$ satisfies all the constraints in $\Phi'$ between
variables of $C$. Indeed, take any two variables $x_1,x_2 \in C$
with a constraint between them. Assume the constraint is created as
a result of some $y\in Y$. Then, since $(L(x_1),L(y)) \in \psi_{x_1
y}$ and $(L(x_2),L(y)) \in \psi_{x_2 y}$, we also have
$(L(x_1),L(x_2)) \in \psi'_{x_1 x_2}$.

It remains to prove the third property. Let $L'$ be a $t$-labeling
to $\Phi'$ and let $C \subseteq X$ be a set of variables of size
$|C| \ge \beta |X|$ with the property that any constraint between
variables of $C$ is satisfied by $L'$. We first define a
$t$-labeling $L''$ to $\Phi$ as follows. For each $x\in X$, we
define $L''(x)=L(x)$. For each $y\in Y$, we define $L''(y) \in
\{1,\ldots,R/d\}$ as the label that maximizes the number of
satisfied constraints between $C$ and $y$. We claim that for each
$y\in Y$, $L''$ satisfies at least $1/t$ of the constraints
between $C$ and $y$. Indeed, for each constraint between $C$ and
$y$ consider the set of labels to $y$ that satisfy it. These sets
are pairwise intersecting since all constraints in $\Phi'$ between
variables of $C$ are satisfied by $L'$. Moreover, since $\Phi$ is
a $d$-to-$1$ label cover, these sets are of size at most $t$.
Claim \ref{clm:pairwiseintersecting} asserts the existence of a
labeling to $y$ that satisfies at least $1/t$ of the constraints
between $C$ and $y$. Since at least $\beta$ of the constraints in
$\Phi$ are incident to $C$, we obtain that $L''$ satisfies at
least $\beta/t$ of the constraints in $\Phi$.

To complete the proof, we define a labeling $L$ to $\Phi$ by
$L(y)=L''(y)$ and $L(x)$ chosen uniformly from $L''(x)$. Since
$|L''(x)| \le t$ for all $x$, the expected number of satisfied
constraints is at least $\beta/t^2$, as required.
\end{proof}

\end{document}